\documentclass{aa}

\usepackage{graphicx}
\usepackage{cuted}

\usepackage[utf8]{inputenc}

\usepackage{txfonts}
\usepackage[dvipsnames]{xcolor}

\usepackage[colorlinks=true,citecolor=blue]{hyperref}

\usepackage{amsmath}
\usepackage{breqn}
\usepackage{tikz}
\usetikzlibrary{quotes,angles}

\newcommand{\kcut}{ k_{\rm cut}}

\newcommand{\cac}{ c_{\rm Ac}}
\newcommand{\uB}{\hat{\textbf{B}}}

\def\vnz{{u_{{\rm n}z}}}
\def\vny{{u_{{\rm n}y}}}
\def\vnx{{u_{{\rm n}x}}}

\def\nunc{{\nu_{\rm nc}}}
\def\nucn{{\nu_{\rm cn}}}
\def\nuni{{\nu_{\rm ni}}}
\def\nunP{{\nu_{\rm nP}}}
\def\nund{{\nu_{\rm nd}}}
\def\sigVL{\langle\sigma v\rangle_{\rm L}}

\def\psin{\psi_{\rm n}}
\def\phic{\phi_{\rm c}}
\def\Atw{\mathcal{A}}
\def\d{{\rm d}}
\def\rhon{\rho_{\rm n}}
\def\rhoc{\rho_{\rm c}}
\def\rhoi{\rho_{\rm i}}
\def\rhoP{\rho_{\rm P}}
\def\rhod{\rho_{\rm d}}
\def\nH{n_{\rm H}}
\def\uc{\textbf{u}_{\rm c}}
\def\un{\textbf{u}_{\rm n}}
\def\pn{p_{\rm n}}
\def\pce{p_{\rm ce}}
\def\gn{g_{\rm n}}
\def\gc{g_{\rm c}}
\def\rhozeron{\rho_{0{\rm n}}}
\def\rhozeroc{\rho_{0{\rm c}}}
\def\nn{n_{\rm n}}
\def\nc{n_{\rm c}}
\def\ni{n_{\rm i}}
\def\nP{n_{\rm P}}
\def\nd{n_{\rm d}}
\def\mn{m_{\rm n}}

\def\mi{m_{\rm i}}
\def\kcut{k_{\rm cut}}
\def\Wcut{\Omega_{\rm cut}}
\def\cAone{c_{\rm Ac1}}
\def\cAtwo{c_{\rm Ac2}}
\def\Onorm{\left(\dfrac{\tilde{\Omega}}{\tilde{\phic}}\right)}
\DeclareMathOperator{\e}{e}

\def\vcz{{u_{{\rm c}z}}}
\def\vcy{{u_{{\rm c}y}}}
\def\vcx{{u_{{\rm c}x}}}

\def \equi#1{\mathrel{\mathop{\kern 0pt\sim}\limits_{#1}}}

\usepackage{subcaption}

\begin{document} 
\makeatletter
\let\linenumbers\relax
\let\endlinenumbers\relax
\makeatother

\title{Impact of the ambipolar diffusion in the structuration of the magnetic Rayleigh Taylor instability with oblique magnetic field}
\titlerunning{Magnetic Rayleigh-Taylor instability at a contact discontinuity}
   
  \author{E. Callies \inst{1}\ ,
                        V. Guillet \inst{1,2}\ ,
                        A. Marcowith \inst{1}\ ,
            Z. Meliani \inst{3}
          \and
            P. Lesaffre\inst{4}\ 
          }

   \institute{Laboratoire Univers et Particules de Montpellier (LUPM) Universit\'e Montpellier, CNRS/IN2P3, CC72, Place Eug\`ene Bataillon, F-34095 Montpellier Cedex5, France
        \and
    Universit\'e Paris-Saclay, CNRS, Institut d’Astrophysique Spatiale, 91405 Orsay, France     
     \and
        Laboratoire Univers et Th\'eories, Observatoire de Paris, Universite PSL, Universite de Paris, CNRS, Meudon, France
        \and
        Laboratoire de Physique de l’\'Ecole Normale Sup\'erieure, ENS, Universit\'e PSL, CNRS, Sorbonne Universit\'e, Universit\'e Paris Cit\'e, 75005, Paris, France
        \\
         \email{edouard.callies@umontpellier.fr} }

 \abstract
   {}
  {We investigate the impact of ambipolar diffusion on the development of the Rayleigh-Taylor  instability (RTI) with an oblique magnetic field in the incompressible limit.}
   {We developed a general bi-fluid framework comprising charges and neutrals with the specific feature of differing gravity between charges and neutrals. We derived the perturbed magnetohydrodynamic (MHD) equations and  obtained an analytic dispersion relation for an oblique magnetic field. The growth rate was then evaluated using a numerical integration of the dispersion relation. In particular, we focussed on the anisotropy in the mode growth induced ambipolar diffusion.}
   {In contrast to the case of a magnetic field within the interface, an oblique magnetic field is much less restrictive with respect to the wavenumbers that can develop; rather than a sharp cut-off, it results in a selection of preferred scales for mode growth. The presence of a second neutral fluid that is insensitive to gravity tends to amplify the anisotropy in the possible direction of the instability development. In particular, we show that this effect is maximal when the coupling is in an intermediate range between full and no charge-neutral coupling, specifically in the region where the ambipolar diffusion is  highest.
}

   {}
\maketitle
\section{Introduction}\label{S:Intro}
The Rayleigh-Taylor instability (RTI)  is one of the main hydrodynamic instability that develops in the interstellar medium (ISM). This instability has, for instance, been thoroughly investigated in the context of pulsar wind nebula; for instance, by \citet{1996ApJ...456..225H, 2004AandA...423..253B}, \citet{2014MNRAS.443..547P}, as well as in the context of supernova remnant expansion (e.g. \citet{1975MNRAS.171..263G} ) and their interaction with molecular clouds \citep{1996ApJ...468L..59J}. \citet{2016MNRAS.462.2256H} presented an overview of analytical and numerical applications of RT in astrophysical systems. The onset of the RT instability has also been invoked in the context of molecular cloud formation \citep{1974AandA....33...73M} and in gas dynamics structure developments in galactic discs \citep{1995ApJ...440..634R}. 

 Generally, the RT instability is triggered when one fluid is resting on top of a less dense fluid and an unstable equilibrium is established.  The resulting dispersion relation is set as follows \citep{Chandrasekhar_1961hhs..book}:
\begin{align}
    \omega^2&=-\Atw gk,\quad \Atw =\dfrac{\rho_2-\rho_1}{\rho_1+\rho_2} \ ,
        \label{eq:relation_HD}
\end{align}
where $g$ is the gravity, $k$ is the wavenumber, and $\Atw$ is the Atwood number, with $\rho_1$ (resp. $\rho_2$) being the density of the light (resp. heavy) fluid. This initial dispersion relation is simple to interpret: whatever the scale, the instability develops and the smaller the scale, the faster the instability develops.

\citet{Chandrasekhar_1961hhs..book} also investigated the impact of the presence of a magnetic field on the development of this instability. In particular, he explored the case of a magnetic field tangent to the interface (i.e. with vanishing obliquity). By deriving the equations within a magnetohydrodynamic (MHD) framework, we can obtain the following dispersion relation (see also the analysis proposed in \citet{1987smh..book.....P},  \citet{1995ApJ...453..332J}):  
\begin{align}
    \omega^2&=-\Atw gk+\dfrac{2(\textbf{B}\cdot\textbf{k})^2}{\mu(\rho_1+\rho_2)} \ ,
    \label{eq:chandrasekhar}
\end{align}
with $\textbf{B}$ the magnetic field, $\textbf{k}$ the wave vector, and $\mu$ the vacuum permeability. The effect of magnetic fields increases at small scales up to the magnetic tension to quench the instability. 

Many theoretical studies have been conducted on the  magnetic RTI (MRTI). The vast majority assume a magnetic field tangent to the discontinuity. This restriction does not allow for a more general behaviour of the magnetic field to be considered. \cite{Vickers} studied the impact of the direction of the magnetic field on the growth rate of the MRTI. The authors proposed a model for an incompressible fluid where the magnetic field has a oblique direction relatively to the interface where the instability develops. \citet{Vickers} concluded that the magnetic field's cutoff is not as abrupt as the one given by Eq.~\ref{eq:chandrasekhar} when the magnetic field leaves the interface. 

In the context of solar physics, the plasma is not totally ionised. \citet{Diaz_2012_2012ApJ...754...41D} developed a model with two fluids (one charged, the other neutral) both compressible, with a magnetic field lying in the interface. The authors investigated the impact of the magnetically constrained charge motion over the development of instability in the neutral fluid. They found that the impact of the charges and their coupling with the neutrals mainly results in an overall slowing down in the development of the instability such that the magnetic field cutoff was almost no longer present (see also \citet{2014AandA...564A..97D}, \citet{2022FrASS...9.9083S}).

In the context of ISM physics, the plasma is known to be magnetised and in the galactic disc to be primarily composed of neutrals (H, H$_2$, or CO depending on the its phase), with a small fraction of charged particles, including ions and grains. ISM is known to often be structured into filaments and is subject to both HD and MHD instabilities and, in particular, the RTI \citep{Jacquet_2011}. In the diffuse ISM, gravity from surrounding stars and self-gravity can be safely ignored. However, the medium can be subjected to radiation pressure exerted by a star on the background medium.  
The impact of this pressure on the medium will be two-fold. On the one hand, the pressure that could push the neutrals is absorbed through the ionisation of the neutrals. Beyond this distance, where the neutrals are ionised, they will no longer be subject to radiation pressure (see chapters  6-7 \citet{2006agna.book.....O}). On the other hand  \citet{10.1063/1.861898} and   \citet{Jacquet_2011} found that a radiative pressure can act similarly to a steady acceleration on the grains, which (through the equivalent principle) will create an gravity field and  trigger the RTI.  

The difference in behaviour between the grains and the neutrals is therefore expected to be significant, as only the grains are subject to both the effective gravity and the magnetic field. With this perspective in mind, we aim to study the MRTI by focusing on three key aspects: the oblique nature of the magnetic field (since an interstellar magnetic field is unlikely to be purely contained within the interface), the fact that gravity affects only the grains, and the inclusion of the neutral fluid, which is dominant in this environment but is subject to neither gravity nor the magnetic field. Our final aim in this paper is to examine to what extent ambipolar diffusion induced by the presence of neutrals can impact the structuring of dust filaments created by the RTI. 

 In Section \ref{S:Model}, we  develop a theoretical model of an incompressible two-fluid system subject to RTI in the presence of an oblique magnetic field, and we derive the dispersion relation that gives the growth rate of the charges. We then highlight in Section \ref{S:Results}, the main differences this relation introduces compared to classical models, namely those of a single-fluid and a two-fluid system, where both fluids experience gravity as well as the impact of the ambipolar diffusion. In Section \ref{S:Appli},  we focus on the application to the interstellar medium, where in our context, only the charges are subject to gravity. We thus study the impact of the magnetic field on the structuring of anisotropy in the development of the RTI to examine the effect of ambipolar diffusion on this anisotropy. We summarise our results and present our conclusions in  Section \ref{S:Concl}.

    \section{Problem formulation and dispersion relation}\label{S:Model}
    \subsection{Plasma configuration}
    
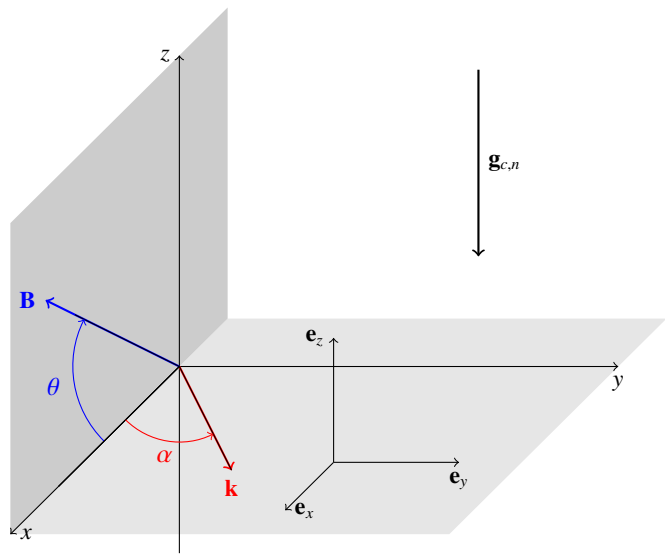
\begin{figure}
    \centering
    \begin{tikzpicture}[scale=1.65]
 
    \fill[gray!20,opacity=0.3] (0,0,-1) -- (3.5,0,-1) -- (3.5,0,3.5) -- (0,0,3.5) -- cycle;
    \fill[gray!40,opacity=0.3] (0,0,-1) -- (0,2.5,-1) -- (0,2.5,3.5) -- (0,0,3.5) -- cycle;

    \draw[->] (0,0,0) -- (3.5,0,0) node[anchor=north] {\footnotesize $y$}; 
    \draw[->] (0,-1.5,0) -- (0,2.5,0) node[anchor=east] {\footnotesize $z$}; 
    \draw[->] (0,0,0) -- (0,0,3.5) node[anchor=west] {\footnotesize $x$};

    \draw[->] (2,0,2) -- (3,0,2) node[anchor=north] {\footnotesize $\textbf{e}_y$}; 
    \draw[->] (2,0,2) -- (2,1,2) node[anchor=east] {\footnotesize $\textbf{e}_z$}; 
    \draw[->] (2,0,2) -- (2,0,3) node[anchor=west] {\footnotesize $\textbf{e}_x$};
    \draw[->, thick] (2,2,-1) -- (2,0.5,-1) node[midway,anchor=west] {\footnotesize $\mathbf{g}_{c,n}$};
   
    \draw[->, thick, red] (0,0,0) -- ({2.5*sin(30)},0,{2.5*cos(30)}) node[anchor=north] {\footnotesize $\mathbf{k}$};

    \draw[->, thick, blue] (0,0,0) -- (0,{3.2*sin(30)},{3.2*cos(30)}) node[anchor=east] {\footnotesize $\mathbf{B}$};

  \draw
     (0,{2.5*sin(30)},{2.5*cos(30)}) coordinate (a) 
    -- (0,0,0) coordinate (b) 
    -- (0,0,2.5) coordinate (c) 
    pic["\textcolor{blue}{\textbf{$\theta$}}", draw=blue, <-, angle eccentricity=1.2, angle radius=1.4cm]
    {angle=a--b--c};

  \draw
    (0,0,2.5) coordinate (a) 
    -- (0,0,0) coordinate (b) 
    --  ({2.5*sin(30)},0,{2.5*cos(30)})coordinate (c) 
    pic["\textcolor{red}{\textbf{$\alpha$}}", draw=red, ->, angle eccentricity=1.2, angle radius=1cm]
    {angle=a--b--c};

\end{tikzpicture}
    \caption{Geometry of the problem. We assume that the $x,y$ plane separates two regions, this is the interface. Each region contains a density $\rho_1$ (for the lower region) and $\rho_2$ (for the upper region) with $\rho_1 < \rho_2$. The entire system is immersed into a magnetic field $\textbf{B}$ lying in the $x,z$ plane, forming an angle $\theta$ with the $x$-axis. We assume that the perturbation develops along the wave vector $\textbf{k}$, which lies in the $x,y$ plane and forms an angle $\alpha$ with the $x$-axis. The charged (resp. neutral) fluid is subject to the gravity field $\textbf{g}_{\rm c}$ (resp. $\textbf{g}_{\rm n}$), directed along $-\textbf{e}_z$.}
    \label{fig:scheme_of_the_framework}
\end{figure}

To keep our approach as general as possible, we chose to consider a bi-fluid system composed of a neutral fluid and a charged fluid. We consider the situation schemed in Fig. \ref{fig:scheme_of_the_framework}. The charged (resp. neutral) species have a volumic mass, $\rhoc$ (resp. $\rhon$). The plane $z = 0$ defines the interface between two regions with different densities: $\rho_{\rm c2}$ (resp. $\rho_{\rm n2}$) for the upper region, and $\rho_{\rm c1}$ (resp. $\rho_{\rm n1}$) for the lower region. We assume that $\rho_{0\rm c2} / \rho_{0\rm c1} = \rho_{0\rm n2} / \rho_{0\rm n1} = d \ge 1$ is constant and uniform. To better encompass all possible physical situations, including real gravity or radiation pressure where the acceleration generated on each species may differ, we assume that each species is subject to a gravitational force directed along $-\textbf{e}_z$, denoted as $\gn$ (for the neutrals) and $\gc$ (for the charges). We also consider that this medium is immersed into a magnetic field $\textbf{B}$, which forms an angle $\theta$ with $\textbf{e}_x$. The instability develops in the interface ported by the axis $\textbf{e}_x$ and $\textbf{e}_y$.

\subsection{Linear analysis}\label{S:LAN}

Our first goal is to port the relations presented in \citet{Vickers} (hereafter V20) to a bi-fluid MHD model.  \citet{Diaz_2012_2012ApJ...754...41D} proposed such kind of analysis but with a magnetic field tangential to the interface. Since we are studying a bi-fluid, we will start with the following equation, using the same framework as \cite{Diaz_2012_2012ApJ...754...41D} : 
\begin{align}
    \rhoc\left(\dfrac{\partial \uc}{\partial t}+\uc\cdot \nabla \uc\right)&=-\nabla \pce+\dfrac{1}{\mu}(\nabla\times\textbf{B})\times \textbf{B}+\rhoc\textbf{g}_{\rm c}\nonumber \\& -\rhoc\nunc (\uc-\un) \ ,\\
    \rhon\left(\dfrac{\partial \un}{\partial t}+\un\cdot \nabla \un\right)&=-\nabla \pn+\rhon\textbf{g}_{\rm n}-\rhon\nucn(\un-\uc) \ ,
\end{align}

where $\uc$  (resp. $\un$) is the velocity of the charged (resp. neutral) fluid, $\pce$ (resp. $\pn$), the pressure of the ion-electron (resp. neutral) fluid, $\nunc$ (resp. $\nucn$),  the frequency of collision for a given charged (resp. neutral) particle with any neutral (resp. charged) particle. Conservation of momentum yields $\rhoc\nunc=\rhon\nucn$.

 The induction equation is expressed as:
\begin{align}
    \dfrac{\partial \textbf{B}}{\partial t}&=\nabla\times(\uc\times \textbf{B}) \ .
\end{align}
To simplify the study, we choose, in a very general case, to assume incompressibility. The justification for the particular case we study in this article is given in Sect. \ref{ss: Context_Appli}. The compressible case is much more complex to investigate. It will deserve a specific future work. The incompressible hypothesis is written in general as :
\begin{align}
    \nabla\cdot\uc&=0 \ , \\
    \nabla\cdot\un&=0 \ .
\end{align}
We assume mass conservation and adiabatic energy equation for all the species and the hydrostatic equilibrium verified for the initial pressures (hereafter, the index 0 refers to equilibrium quantities):
\begin{align}
     \dfrac{\partial p_{0{\rm ce}}}{\partial z}&=-\gc\rhozeroc \ ,\\
    \dfrac{\partial p_{0{\rm n}}}{\partial z}&=-\gn\rhozeron \ .
\end{align}
With these considerations we obtain our system of basic equations:
\begin{align}
     \rhoc\left(\dfrac{\partial \uc}{\partial t}+\uc\cdot \nabla \uc\right)&=-\nabla \pce+\dfrac{1}{\mu}(\nabla\times\textbf{B})\times \textbf{B}+\rhoc\textbf{g}_{\rm c}\nonumber \\& -\rhoc\nunc(\uc-\un) \ ,\\
    \rhon\left(\dfrac{\partial \un}{\partial t}+\un\cdot \nabla \un\right)&=-\nabla \pn+\rhon\textbf{g}_{\rm n}-\rhon\nucn(\un-\uc) \ ,\\
     \dfrac{\partial \pce}{\partial t}+\uc\cdot \nabla \pce&=0\ ,\\
     \dfrac{\partial \pn}{\partial t}+\un\cdot \nabla \pn&=0\ ,\\
    \dfrac{\partial \rhoc}{\partial t}+\uc\cdot\nabla\rhoc&=0 \ , \\
    \dfrac{\partial \rhon}{\partial t}+\un\cdot\nabla\rhon&=0 \ ,\\
    \dfrac{\partial \textbf{B}}{\partial t}&=\nabla\times(\uc\times \textbf{B}) \ .
\end{align}

\subsubsection{System linearisation}
To find the dispersion relation, we linearised the previous equations with respect to perturbations written in their Fourier form as $\e^{i(k_xx+k_yy-\omega t)}$ in each region separately. Since our study is aimed at studying the growth rate of the instability, we introduced $\Omega=-i\omega$, a positive real that defines our perturbation as $\e^{i(k_xx+k_yy)}\e^{\Omega t}$.  We  connected the solution of each region with our boundary conditions. We called $\textbf{b}$ the perturbation of the magnetic field and we normalised the magnetic field to $B_0$. Hence, we have the notations $\vec{\hat{b}}= \vec{b}/B_0, \vec{\hat{B}}= \vec{B}/B_0$. We retained all terms to the first order. With these considerations,  we can obtain the following system:

\begin{align}
    \Omega\rhozeroc \vcx&=-ik_x\pce-\rhozeroc\nunc(\vcx-\vnx)\nonumber \label{Eq:VCX}\\ &+\rhozeroc\cac^2\sin\theta\,(\partial_z\hat{b}_x-ik_x\hat{b}_z) \ ,\\
     \Omega\rhozeroc \vcy&=-ik_y\pce-\rhozeroc\nunc(\vcy-\vny)\nonumber \\
    &+\rhozeroc\cac^2\left[\sin\theta\,(\partial_z\hat{b}_y-ik_y\hat{b}_z)\right.\nonumber\\
    &+\left.\cos\theta\,(ik_x\hat{b}_y-ik_y\hat{b}_x)\right]  \ ,\label{Eq:VCY}\\
    \Omega\rhozeroc\vcz&=-\dfrac{\partial \pce}{\partial z}-\rhozeroc\nunc(\vcz-\vnz)\nonumber\\ &+\rhozeroc\cac^2\cos\theta\,(ik_x\hat{b}_z-\partial_z\hat{b}_x)-\rhoc\gc \label{Eq:VCZ} \ ,\\
     \Omega\rhozeron \vnx&= -ik_x\pn -\rhozeron\nucn(\vnx-\vcx) \label{Eq:VNX} \ ,\\
   \Omega\rhozeron \vny&=-ik_y\pn-\rhozeron\nucn(\vny-\vcy) \label{Eq:VNY} \ ,\\
   \Omega \rhozeron\vnz &=- \dfrac{\partial \pn}{\partial z} -\rhozeron\nucn(\vnz-\vcz)-\rhon \gn \label{Eq:VNZ} \ ,\\
   \Omega \pce&=-\vcz\partial_z p_{0{\rm ce}} \ ,\\
   \Omega \pn&=-\vnz\partial_z p_{0{\rm n}} \ ,\\
   \Omega\rhoc&=-\vcz\partial_z \rhozeroc \ ,\\
   \Omega\rhon&=-\vnz\partial_z \rhozeron \ ,\\
   \Omega \hat{\textbf{b}}&=\hat{\textbf{B}}\cdot \nabla\uc \ ,
\end{align}
with 
\begin{equation}
    \cac\equiv\sqrt{\dfrac{B_0^2}{\mu\rhozeroc}} \ ,
\end{equation}
the Alfven velocity in the charged fluid. At this point, we can express the components of the neutral velocity in Eqs. (\ref{Eq:VNX}-\ref{Eq:VNZ}) as a function of the components of the charges velocity and subsituted them into Eqs. (\ref{Eq:VCX}-\ref{Eq:VCZ}). Then, deriving according to time, we can replace the expressions of the magnetic components and we obtain 
\begin{align}
    \dfrac{\Omega^2}{\phic^2}\vcx&=   -\dfrac{ik_x\textbf{$\Omega$}}{\rhozeroc}\left(\pce+\pn \psin\right)\nonumber\\ &+\cac^2(\uB\cdot\nabla)\left[\sin\theta\,(\partial_z\vcx-ik_x\vcz)\right]\label{eq:vix_final} \ ,\\
    \dfrac{\Omega^2}{\phic^2}\vcy&=-\dfrac{ik_y\textbf{$\Omega$}}{\rhozeroc}\left(\pce+\pn \psin\right)\nonumber\\
    &+\cac^2(\uB\cdot\nabla)\left[\sin\theta\,(\partial_z\vcy-ik_y\vcz)\right.\nonumber\\
    &+\left.\cos\theta\,(ik_x\vcy-ik_y\vcx)\right]\label{eq:viy_final} \ ,\\
    \dfrac{\Omega^2}{\phic^2}\vcz&=-\dfrac{\textbf{$\Omega$}}{\rhozeroc}\dfrac{\partial}{\partial z}\left(\pce+ \pn\psin\right)\nonumber\\
    &+\cac^2(\uB\cdot\nabla)\left[\cos\theta\,(ik_x\vcz-\partial_z\vcx)\right]\nonumber\\
    &+\dfrac{\partial_z\rhozeron}{\rhozeron}\psin \gn\vnz+\dfrac{\partial_z\rhozeroc}{\rhozeroc}\gc\vcz\label{eq:viz_final} \ ,
\end{align}

 where we have introduced the adimensional quantities functions of the growth rate $\Omega$ and the collisional frequencies $\nunc$ and $\nucn$:
\begin{align}
    \psin&\equiv\dfrac{\nucn}{\Omega+\nucn}\ , \\
    \phic&\equiv\sqrt{\dfrac{\Omega+\nucn}{\Omega+\nucn+\nunc}}\ .
\end{align}
To eliminate the pressure terms, we  calculated $\partial_z(\ref{eq:vix_final})-ik_x\times (\ref{eq:viz_final})$ and $k_y(\ref{eq:vix_final})-k_x\times (\ref{eq:viy_final}).$  We obtained the following system:
\begin{align}
     \left[\cac^2\phic^2(\uB\cdot\nabla)^2-\Omega^2\right]\left(\partial_z\vcx -ik_x\vcz\right)& =\nonumber\\
      ik_x\phic^{2}\left[ \dfrac{\partial_z\rhozeron}{\rhozeron}\psin \gn\vnz+\dfrac{\partial_z\rhozeroc}{\rhozeroc}\gc\vcz\right]+\Omega^2\left[\dfrac{\partial_z\rhozeroc}{\rhozeroc}\vcx\right]& \ ,\\
    \left[\cac^2\phic^2(\uB\cdot\nabla)^2-\Omega^2\right]\left(-i\dfrac{k^2}{k_x}\vcx-\partial_z\vcz\right)& = 0  \ .
\end{align}
comparing our equation and the one obtained in V20 (their Eq. 13), we can define the same function $f$ by replacing $v_{\rm A}^2$ by $\cac^2\phic^2$: 

\begin{equation}
  f\equiv\cac^2\phic^2(\hat{\textbf{B}}\cdot \nabla)^2-\Omega^2  \ .
\end{equation}

At this stage, we can adopt the same assumption as V20, namely, $\partial_z\rhozeroc = \rhozeroc'= 0$ and $\partial_z\rhozeron = \rhozeron'= 0$. This assumption is motivated by two main aspects. First, in our model, we a priori assume that there is a uniform density on either side of the interface and we treat each side of the interface separately. Thus, even if we have a discontinuity of the density across the interface, the density of each side is constant. This simplifies the mathematical treatment and aligns with the physical intuition that the system is initially homogeneous at large scales. Second, as argued bin V20, the instability operates on a length scale much smaller than the characteristic scale over which the density varies. In other words, the density profile does not change significantly over the small spatial regions where the instability develops. This separation of scales justifies neglecting the spatial variation of the density when analysing the instability. With this assumption in place, we recovered the same differential equation for $f$ as V20. This equivalence further validates our approach and confirms that the underlying physics of the instability remains consistent under these assumptions. The charge velocity hence verifies the ordinary differential equation
\begin{equation}
    \left(k^2-\dfrac{\partial^2}{\partial z^2}\right)f(\vcz)=0  \ .
\end{equation}
By developing this equation, we can obtain a linear differential equation in $\vcz$. The roots of its characteristic equation are then:
\begin{equation}
    S=\left\{\pm k, i\left(\dfrac{k_x}{\tan\theta}\right)\pm\dfrac{\Omega}{\sin\theta\,\cac\phic}\right\}  \ .
\end{equation}

From this point, we would generally require a complex growth rate, on the one hand, but also call for the phenomenon to be evanescent, as the growth starts at $z=0$ and $t=0$. With this in mind, we can explicitly write the expression for the $z$-component of the velocity as:
\begin{equation}
      \vcz=
    \begin{cases}
        A_1\e^{kz}+B_1\e^{m_{1-}z}, z<0,\\
        A_2\e^{-kz}+B_2\e^{m_{2+}z}, z>0.
    \end{cases} \ 
\end{equation}
We see that the $x$-component of the velocity is also a linear combination of the same exponential. We have:
\begin{equation}
     \vcx=\begin{cases}
        i\dfrac{k_x}{k}A_1\e^{kz}+C_1\e^{m_{1-}z}, z<0,\\
        -i\dfrac{k_x}{k}A_2\e^{-kz}+C_2\e^{m_{2+}z}, z>0.
    \end{cases} \ 
\end{equation}
To fully characterise our system, we also need to derive expressions for the $x$ and $z$ components of the velocity of the neutrals. By using the equation of motion for the neutrals (Eqs. \ref{Eq:VNX}-\ref{Eq:VNZ}) and applying the same transformation as for the charges equations we obtain, under the assumption that $\rhozeron'=0$, the following relation:
\begin{equation}
     \begin{pmatrix}
        \partial_z \vnx-ik_x\vnz\\
        -ik^2\vnx-k_x\partial_z\vnz
    \end{pmatrix}=\psin \begin{pmatrix}
        \partial_z \vcx-ik_x\vcz\\
        -ik^2\vcx-k_x\partial_z\vcz
    \end{pmatrix} \ .
\end{equation}
With this relation, it becomes clear that the $x$ and $z$ components of the velocity of the neutrals are also a combination of the two same exponential as the $x$ and $z$ components of the velocity of the charges. More precisely, we can write:
\begin{align}
     \vnz&=
    \begin{cases}
        D_1\e^{kz}+\psin B_1\e^{m_{1-}z}, z<0,\\
        D_2\e^{-kz}+\psin B_2\e^{m_{2+}z}, z>0,
    \end{cases} \ \\
    \vnx&=
    \begin{cases}
        i\dfrac{k_x}{k}D_1\e^{kz}+\psin C_1\e^{m_{1-}z}, z<0,\\
        -i\dfrac{k_x}{k}D_2\e^{-kz}+\psin C_2\e^{m_{2+}z}, z>0.
    \end{cases} \ 
\end{align}

\subsubsection{Boundary conditions}
Given the form of the $x$ and $z$ velocity components for the charges and neutrals, we required eight boundary conditions to determine our dispersion relation. The first condition is meant to consider the continuity of $\uc$. We can also add the incompressibility and the continuity of the magnetic field at the interface to obtain five boundary conditions:
\begin{align}
    \left[\vcz\right]&=0 \label{eq:vcz_continuity} \ ,\\
    \left[\vcx\right]&=0 \label{eq:vcx_continuity} \ ,\\
    \left[\dfrac{\partial \vcz}{\partial z}\right]&=0 \ ,\\
    \left[\dfrac{\partial^2 \vcz}{\partial z^2}\right]&=0 \ ,\\
    \left[\dfrac{\partial \vcx}{\partial z}\right]&=0 \ ,
\end{align}
where  $\left[X\right]$ stands for the jump of the quantity $X$ across the interface. We used the energy equations  for charges assuming its continuity, which gives :
\begin{align}
      \left[\gc\rhoc\vcz-\dfrac{\partial \pce}{\partial t}\right]&=0 \ .
\end{align}
To obtain the eight mandatory boundary conditions, we added the continuity of the $z$ component of $\un$ and the continuity of the energy equation,
\begin{align}
    \left[\vnz\right]&=0 \ ,\\
     \left[\gn\rhon\vnz-\dfrac{\partial \pn}{\partial t}\right]&=0 \ .\label{eq:presssure_n_continuity}
\end{align}
To obtain a proper matrix for the boundary conditions, we replaced the expression of the pressure by using the equation of motion on the $x$ component for each species, which gives:
\begin{align}
    [\rhozeroc\left(\phic^2 
   \Omega(\nunc+\Omega)-\cac^2\phic^2\sin^2\theta\partial_z^2\right)\vcx&\nonumber\\
   +\phic^2\rhozeroc\Omega\nunc\vnx+ik_x \phic^2\gc\rhozeroc\vcz]&=0 \ ,\\
\left[
   \Omega\rhozeron(\nucn+\Omega)\vnx+ \rhozeron\Omega\nucn\vcx +ik_x\gn\rhozeron\vnz\right]&=0 \ .
\end{align}

Using the expression of $\vcx,\vcz,\vnx, \vnz$, we can express the boundary conditions in a matrix form with each line corresponding respectively to Eqs. \ref{eq:vcz_continuity}-\ref{eq:presssure_n_continuity}
as follows:\begin{align}
     \begin{pmatrix}
        1&1&0&-1&-1&0&0&0\\
        ik_x/k&0&1&ik_x/k&0&-1&0&0\\
        k&m_{1-}&0&k&-m_{2+}&0&0&0\\
        k^2&m_{1-}^2&0&-k^2&-m_{2+}^2&0&0&0\\
        ik_x&0&m_{1-}&-ik_x&0&-m_{2+}&0&0\\
        \alpha_{1{\rm c}}&\beta_{1{\rm c}}&\gamma_{1{\rm c}}&-\alpha_{2{\rm c}}&-\beta_{2{\rm c}}&-\gamma_{2{\rm c}}& \delta_{1{\rm c}}&-\delta_{2{\rm c}}\\
        0&\psin&0&0&-\psin&0&1&-1\\
       \alpha_{1{\rm n}}&\beta_{1{\rm n}}&0&-\alpha_{2{\rm n}}&-\beta_{2{\rm n}}&0& \delta_{1{\rm n}}&-\delta_{2{\rm n}}\\
    \end{pmatrix} \begin{pmatrix}
             A_1\\
             B_1\\
             C_1\\
             A_2\\
             B_2\\
             C_2\\
             D_1\\
             D_2\\
     \end{pmatrix}&=&0\nonumber \ ,\\
\end{align}

with
\begin{align}
    \alpha_{1{\rm c}}&\equiv k_x\phic^2\left[\gc-\left(\cAone^2k\sin^2\theta-\dfrac{\Omega\left(\nunc+\Omega\right)}{k}\right)\right] \ ,\\
    \alpha_{2{\rm c}}&\equiv d^{-1}k_x\phic^2\left[ \gc+\left(d\cAone^2k\sin^2\theta-\dfrac{\Omega\left(\nunc+\Omega\right)}{k}\right)\right] \ ,\\
    \beta_{1{\rm c}}&\equiv k_x\phic^2\gc \ ,\\
    \beta_{2{\rm c}}&\equiv d^{-1}k_x\phic^2\gc \ ,\\
    \gamma_{1{\rm c}}&\equiv i\left(\phic^2\cAone^2m_{1-}^2\sin^2\theta-\Omega^2\right) \ ,\\
    \gamma_{2{\rm c}}&\equiv d^{-1}i\left(d\phic^2\cAone^2m_{2+}^2\sin^2\theta-\Omega^2\right) \ ,\\
    \delta_{1{\rm c}}&\equiv -k_x\phic^2\dfrac{\Omega\nunc}{k} \ ,\\
    \delta_{2{\rm c}}&\equiv d^{-1}k_x\phic^2\dfrac{\Omega\nunc}{k} \ ,\\
    \alpha_{1{\rm n}}&\equiv -k_x\dfrac{\Omega\nucn}{k} \ ,\\
    \alpha_{2{\rm n}}&\equiv d^{-1}k_x\dfrac{\Omega\nucn}{k} \ ,\\
    \beta_{1{\rm n}}&\equiv k_x\psin \gn \ ,\\
    \beta_{2{\rm n}}&\equiv d^{-1}k_x\psin \gn \ ,\\
    \delta_{1{\rm n}}&\equiv k_x\left(\gn+\dfrac{\Omega(\nucn+\Omega)}{k}\right) \ ,\\
    \delta_{2{\rm n}}&\equiv d^{-1}k_x\left(\gn-\dfrac{\Omega(\nucn+\Omega)}{k}\right) \ .
\end{align}
We defined $\cAone\equiv \sqrt{B_0^2/(\mu\rho_{0\rm c1})}$ as the Alfven velocity in the medium one (the light medium). We replace the analogous $\cAtwo$ by $d^{1/2}\cAone$. We can see that if we suppose that the two fluids are fully decoupled (i.e. $\nucn=\nunc=0$) we will obtain a block matrix with the first 6 rows and column giving the matrix obtained by V20 and the last two  giving the classic HD dispersion relation: $\Omega_{\rm n}^2=\Atw\gn k$. The next step is to further develop the determinant and after some algebraic manipulations, we obtain:
\begin{equation}
     \Delta_{\rm bi}(k-m_{1-})(k+m_{2+})(m_{1-}+m_{2+})=0 \ ,
\end{equation}
where
\begin{align}
     \Delta_{\rm bi}&\equiv   \Delta_{\rm MHD}\left(1-\dfrac{\Omega(\nucn+\Omega)}{\Atw\gn k}\right)\\
     &+2k^2\phic^3 \gn\cAone\dfrac{\nunc\nucn}{(\nucn+\Omega)^2}  \sin\theta\frac{1 - d}{1 + d^{-1/2}}\nonumber\\ 
     &+k\phic^2 \gn\dfrac{\nunc\nucn}{(\nucn+\Omega)^2}(1 - d)\Omega \nonumber \ ,
\end{align}
and
\begin{align}
    \Delta_{\rm MHD} &\equiv(1+d)\Omega^3+2kd^{1/2}\phic \cAone\sin\theta\,(1+d^{1/2})\Omega^2
   \\
&+\phic^2\left[-k\tilde{g}(1-d)+2k^2d\cAone^2(1-\sin^2\alpha\cos^2\theta)\right]\Omega\nonumber\\
&-2k^2\phic^3 \cAone\tilde{g}\sin\theta\,\dfrac{d^{1/2}(1-d)}{1+d^{1/2}}\nonumber \ ,
\end{align}
with 
\begin{equation}
    \tilde{g} \equiv \gc+\gn \frac{\nunc\nucn}{(\nucn+\Omega)^2}\,.
\end{equation}

The discriminant $\Delta_{\rm MHD}$ corresponds to the expression of $S$ given by V20 in their Eq. 26, with a correction\footnote{Vickers writes the $\omega^3$ term as $-(d^{1/2}+1)(1-d)\omega^3$. The correct expression is $-(d^{1/2}+1)(1+d)\omega^3$. This typo does not impact their results.}. The effect of the second fluid on this expression could be summed up by the substitution of $g\rightarrow \phic^2\tilde{g}$ and $v_{\rm A}\rightarrow\phic \cAone$.
\section{Results}\label{S:Results}

\subsection{Context and normalisation}
We start by defining $\Omega_{\rm MHD}$ as the solution to the dispersion relation obtained by setting $\nucn=0$, namely, the solution obtained in V20. We use as a normalisation for  wavenumbers:

\begin{equation}
    \kcut\equiv\dfrac{(1-d)\gc}{2d\cAone^2} \ .
    \label{eq:kcut}
\end{equation}

This represents the maximum wavenumber value for which the instability can develop with $\vec k$ along $\vec B$ when the magnetic field is tangent ($\theta = 0^o$), namely, solving $\omega=0$ for $\alpha=0$  in Eq. \ref{eq:chandrasekhar} .
We define the associated pulsation:
\begin{equation}
\Wcut\equiv\sqrt{\Atw\gc \kcut} = \dfrac{(1-d)\gc}{\sqrt{2d(1+d)}\,\cAone} \ .
    \label{eq:wcut}
\end{equation}
All frequencies, if not specified otherwise, have been normalised by this reference frequency. Specifically, we used normalised coupling parameter $\nunc/\Wcut$. Physically, this normalised quantity is a way to compare the effectiveness of the coupling to the characteristic timescale for the growth of the magnetic RT instability.  In order to solve the dispersion relation, we used Python's \texttt{fsolve} routine to extract the localisation of the roots. We then selected the one(s) with a positive real part.

\subsection{Gravity on the two species}

\begin{figure}
    \centering
     \includegraphics[width=\linewidth]{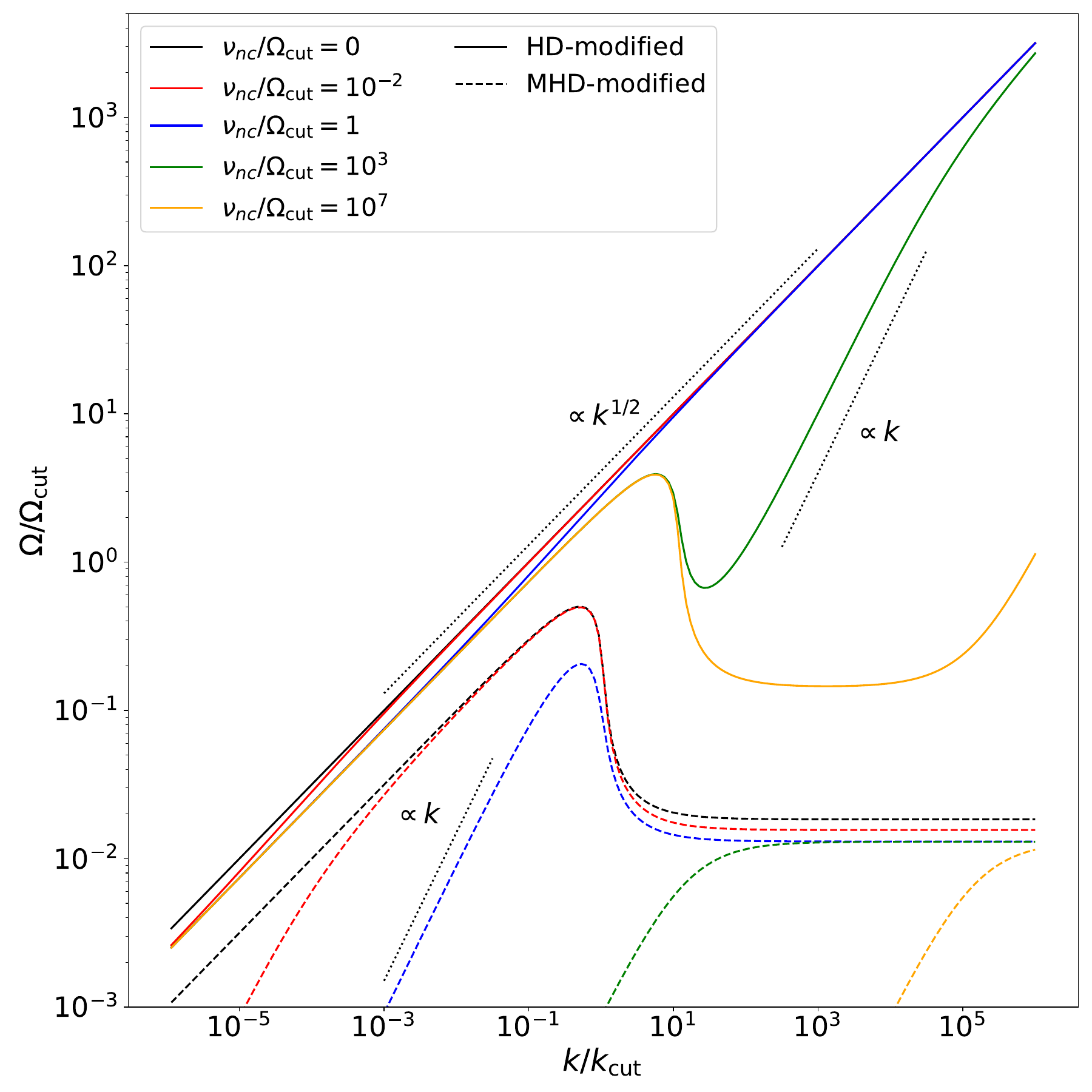}
    \caption{Solutions to the dispersion relation, normalised by$\Wcut$, for various values of the coupling parameter $\nunc/\Wcut$:  $0$ (black lines), $10^{-2}$ (red lines) , $1$ (blue lines), $10^{3}$ (green lines), and $10^{7}$ (yellow lines). We plot the two modes obtained: HD modified (plain lines) MHD modified (dashed lines). We set $\theta=1^\circ$, $\alpha=0^\circ$, $d=0.25$, $\gn=10\gc$, $\rhoc=\rhon$. }
    \label{fig:gn=gi_theta=0}
\end{figure}

The dispersion relation  $\Delta_{\rm bi}(\Omega)=0$ has two solutions for $\gn\ne 0$ and $\gc\ne0$. In this section, to facilitate identifying in figures the different modes destabilised by the action of the gravity fields associated to $\gn$ and $\gc$ we set $\gn=10\gc$. We also set $\rhon=\rhoc$ so that $\nucn=\nunc$. We show the solution to the dispersion relation in Fig. \ref{fig:gn=gi_theta=0}. We normalise the growth rate to $\Wcut$. In the case of a vanishing coupling ($\nunc = 0$), there are two solutions to the dispersion relation. The first one corresponds to the growth of the instability in the neutral fluid ("HD" mode); it scales as $\sqrt{\Atw \gn k}$. The second is the one that is obtained by solving the V20 dispersion relation ("MHD" mode). As soon as ion-neutral coupling is activated, the two modes are modified ("HD-modified" and "MHD-modified" modes, respectively) and cannot directly be identified to development of the instability in the neutral or charged fluid, respectively. \\

\subsubsection{Behaviour of the MHD-modified mode}
Without coupling ($\nunc = 0$), the behaviour of the solution for the MHD-modified mode corresponds to the behaviour described by V20: since the magnetic field is not purely tangent to the interface ($\theta=1^\circ$), the cut-off effect at $\kcut$ disappears. The growth rate reaches a maximum before tending towards an asymptote given by equation 29 in V20.  With coupling, the behaviour is different. The asymptote for large $k$ is still present, but its value varies. By expanding our dispersion relation and retaining only the leading orders in $k$, we find that:

\begin{align}
     \dfrac{\Omega}{\phic}(k\to\infty)&\longrightarrow\Omega_{\rm MHD,k_\infty} \ ,\label{eq:Asymptotic_Omega_value} 
\end{align}
where
\begin{align}
     \Omega_{\rm MHD,k_\infty}&\equiv
     \dfrac{\gc}{\cAone}\times\dfrac{ (1 - d)\sin \theta 
}{d^{1/2}\left(1 + d^{1/2}\right) 
\left( 1 - \sin^2 \alpha \cos^2 \theta \right)} \ ,
\end{align}
is the asymptotic value given by Eq. 29 of V20. Therefore in a general case, we can find the asymptotic value for large $k$ by solving this third-order equation:
\begin{align}
    \Omega^3+(\nunc+\nucn)\Omega^2-\Omega_{\rm MHD,k_\infty}^2\Omega - \nucn\Omega_{\rm MHD,k_\infty}^2 &= 0 \ .
\end{align}
Overall, this equation has no simple analytical solutions. However, we can calculate the solution for extreme value of $\nunc$. Indeed, taking the limit $\nunc/\Omega\ll 1$, $\phic$ tends towards 1 and $\Omega$ tends towards $\Omega_{\rm MHD,k_\infty}$. At the opposite end, for $\nunc/\Omega\gg 1$, $\phic$ tends towards $\sqrt{\rhoc/(\rhoc+\rhon)}$,  which explains why in our case of $\rhoc=\rhon$, the asymptote is reduced by a factor $\sqrt{2}$ with respect to V20. \\

 In the general case, the behaviour depends on whether $\nunc \ll \Omega$ or $\nunc \gg \Omega$. Indeed, when $\nunc \ll \Omega$, we observe that the growth rate corresponds almost exactly to the growth rate in the uncoupled case. For instance, we can plot a horizontal line for $\nunc/\Wcut = 10^{-2}$ (red dashed curves) at $\Omega/\Wcut = 10^{-2}$. Above this horizontal line, the growth rate corresponds almost to that of the uncoupled case. At small $k$, the growth rate transitions to a linear growth in ${k}$. This characteristic effect can be understood from a physical point of view. For this, we compare, on the one hand, the characteristic growth time of the instability and, on the other, the characteristic coupling time. When the former is much shorter than the latter, the instability develops too fast for the coupling to have any significant impact. When the latter is much longer than the former, the instability can only occur by considering a strong coupling. As $\nunc$ increases, the range of $k$ such that $\Omega > \nunc$ shrinks and the entire growth rate eventually transitions to a linear growth in $k$ before reaching the asymptote described above. For an infinite value of the coupling parameter, the MHD-modified mode completely disappears.

\subsubsection{Behaviour of the HD-modified mode}

Without coupling, the solution for the HD-modified mode behaves as predicted by Eq. \ref{eq:relation_HD}, namely, a scaling as $\sqrt{\Atw\gn k}$. We first considered small values for the coupling parameter ($\nucn\ll\Wcut$). At the largest scales ($k\ll \kcut$), the charged and neutral fluids are perfectly coupled and are subject to the effective gravity field $g_{\rm bf}$ defined as:
\begin{equation}
    g_{\rm bi}\equiv \dfrac{\rhoc\gc+\rhon\gn}{\rhoc+\rhon} \ ,
\end{equation}
which is the weighted average of the gravity experienced by the two species and ultimately assumes the fluids are perfectly coupled. At smaller scales ( $k\gg \kcut$), as soon as $\Omega > \nucn$, the neutral fluid decouples from the charged fluid and we recover the trend followed by the neutral gas in the uncoupled case. Now, at large values of the coupling parameter ($\nucn\gg\Wcut$), the two fluids are strongly coupled at all $k$ and follow the trend of the charged fluid in the uncoupled case, but with an effective gravity field $g_{\rm bi}$ and an effective $k_{\rm bi,cut}$:
\begin{equation}
    k_{\rm bi,cut}\equiv\kcut\dfrac{g_{\rm bi}}{\gc} \ .
\end{equation}

For the largest value plotted, we observe a dip given by :
\begin{equation}
     \Omega=\sqrt{\dfrac{\rhoc+\rhon}{\rhoc}}\dfrac{g_{\rm bi}}{\cAone}\times\dfrac{ (1 - d)\sin \theta 
}{d^{1/2}\left(1 + d^{1/2}\right) 
\left( 1 - \sin^2 \alpha \cos^2 \theta \right)} \ .
\end{equation}

 As $\nucn$ increases, the range of $k$ such that $\Omega$ follows this equation expands, so that when the coupling becomes infinite, we can obtain an asymptote for large values of $k$.

 \subsubsection{Discussion}
This study bears similarities to the one conducted by \citet{Diaz_2012_2012ApJ...754...41D},  investigating MRTI with two fluids, along with the effect of magnetic field, but having it assigned to the interface. The authors also obtained two modes: a 'neutral' one and an 'ionic' one (see Fig.5 of their paper).  As in the present study,  the presence of coupling with the charged fluid does not completely suppress the instability in the neutral mode (which we refer here as to the 'HD' mode). Just as we have found, the presence of coupling with the charged fluid does not completely suppress the instability in the neutral mode (i.e. our 'HD' mode). Moreover, as the coupling increases, these authors  also observed that the HD mode tends to weaken, similarly to the behaviour shown by the yellow curve in Fig. \ref{fig:gn=gi_theta=0}. It is less straightforward to compare the case of charge modes in Diaz’s work with the MHD mode here, since the orientation of the magnetic field differs between the two cases. However, it is worth noting that they also observe an attenuation of the charge mode as the coupling increases.

\subsection{Case of $\gn=0$}
\subsubsection{Dispersion relation}
If we assume that the radiative forcing is only impacting the charges, we can set $\gn=0$. With some algebraic manipulation, the matrix of the boundary conditions can  be expressed as:
\begin{eqnarray}
\begin{pmatrix}
        1&1&0&-1&-1&0\\
        ik_x/k&0&1&ik_x/k&0&-1\\
        k&m_{1-}&0&k&-m_{2+}&0\\
        k^2&m_{1-}^2&0&-k^2&-m_{2+}^2&0\\
        ik_x&0&m_{1-}&-ik_x&0&-m_{2+}\\
       \tilde{\alpha}_1&\tilde{\beta}_1&\tilde{\gamma}_1&   -\tilde{\alpha}_2&-\tilde{\beta}_2&-\tilde{\gamma}_2
    \end{pmatrix}
    \begin{pmatrix}
             A_1\\
             B_1\\
             C_1\\
             A_2\\
             B_2\\
             C_2\\
     \end{pmatrix}&=&0\ ,
\end{eqnarray}
where the greek tilde letters represents the same coefficients as the one of the V20 matrix with the difference that $V_A$ is here $\cac\phic$, $g$ is $\gc\phic^2$ and the sign in $m_{2+}$ and $m_{1-}$ are inverted, since we choose to study $\Omega=-i\omega$ rather than $\omega$:
\begin{align}
     \tilde{\alpha}_1 &\equiv  k_x \phic^2\gc-\dfrac{k_x}{k}\left(\phic^2\cAone^2k^2\sin^2\theta -\Omega^2\right),\\
 \tilde{\alpha}_2 &\equiv  d^{-1}\left(k_x \phic^2\gc+\dfrac{k_x}{k}\left(\phic^2 d \cAone^2k^2\sin^2\theta -\Omega^2\right)\right) 
 ,\\
 \tilde{\beta}_1 &\equiv  k_x\phic^2\gc,\\
 \tilde{\beta}_2 &\equiv  d^{-1} k_x\phic^2\gc,\\
 \tilde{\gamma}_1 &\equiv  i\left(\phic^2 \cAone^2m_{1-}^2\sin^2\theta -\Omega^2\right),\\
 \tilde{\gamma}_2 &\equiv   id^{-1}\left(\phic^2d \cAone^2m_{2+}^2\sin^2\theta -\Omega^2\right).
\end{align}

Therefore the dispersion relation is similar to the one derived V20, but with the substitution mentioned above. Thus,  we have to solve:
\begin{align}
\Delta_{\rm MHD} = 0, \text{ with } \tilde{g}\equiv \gc\,.
        \label{eq:dispersion_gn0}
\end{align}

This determinant is slightly more complex than the one obtained by V20 since both $\Tilde{g}$ and $\phic$ depend upon $\Omega$. In order to centralise the discussion on the coupling effect, we decide to use normalised quantity $k/\kcut$ and $\Omega/\Omega_{\rm cut}$. With those quantities we can rewrite Eq.~\eqref{eq:dispersion_gn0} as:
\begin{align}
0&=\Onorm^3+\Onorm^2\tilde{k}\sqrt{2}\,a\sin\theta \nonumber\\
&+\Onorm\left(-\tilde{k}+\tilde{k}^2\left(1-\sin^2\alpha\cos^2\theta\right)\right)-\tilde{k}^2\dfrac{\sqrt{2}\sin\theta}{a} \ , 
\end{align}
using the notation $a\equiv(1+d^{1/2})/(1+d)^{1/2}$.

\begin{figure}
    \centering
    \includegraphics[width=\linewidth]{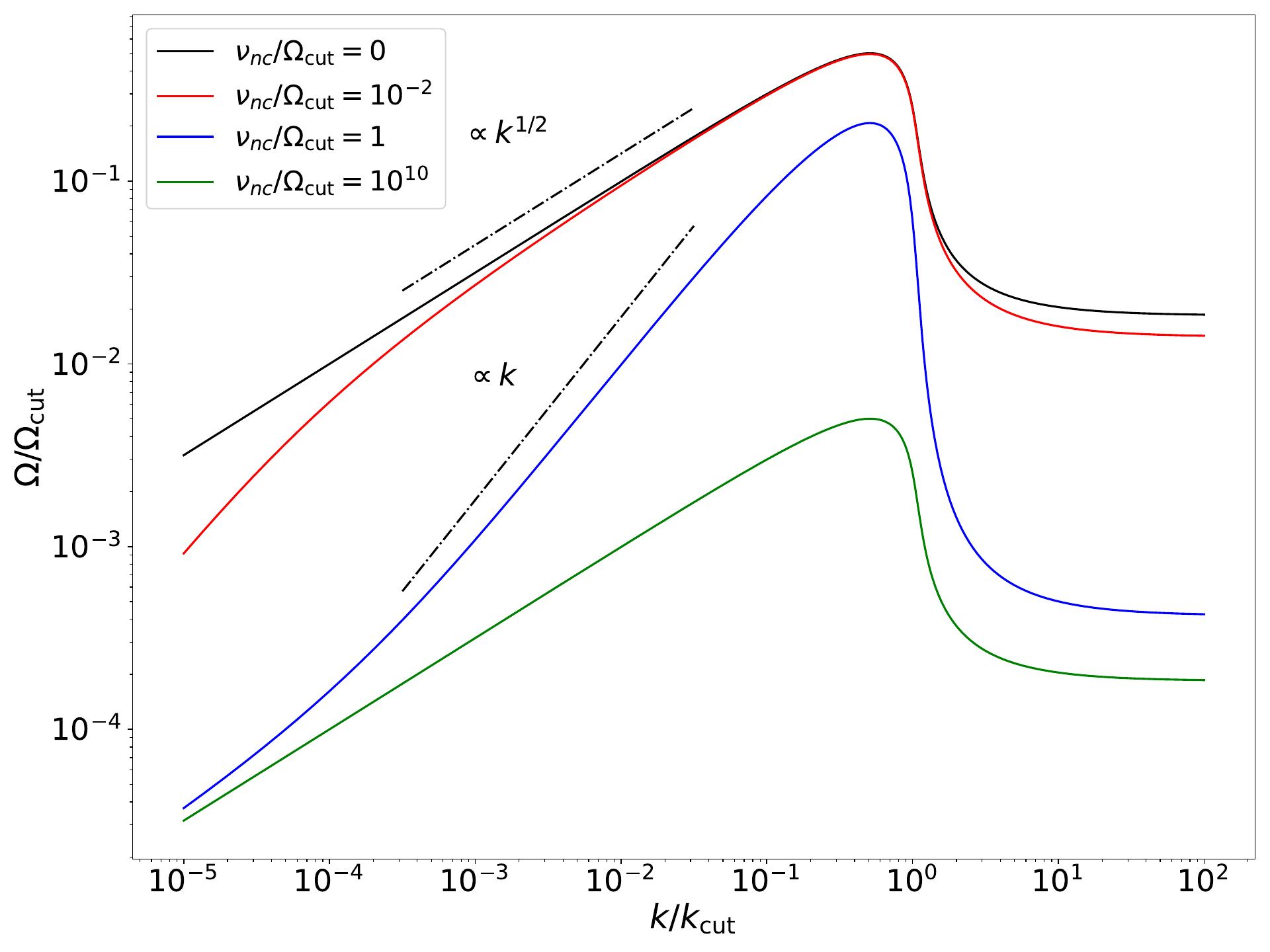}
    \caption{Solutions to the dispersion relation for various values of the coupling parameters. The solutions are normalised by $\Wcut$. We set $\gn=0$, $\theta=1^\circ$, $\alpha=0^\circ$, $d=0.25$, and $\rho_{0c}/\rho_{0n}=10^{-4}$.}
    \label{fig:omega_vs_k_gn0}
\end{figure}

We plot the solution of the dispersion relation normalised by $\Wcut$ in Fig. \ref{fig:omega_vs_k_gn0} for various values of the coupling parameter. To see the impact of the friction of the neutral fluid, we set $\rhozeroc/\rhozeron=10^{-4}$. Without any coupling, we obtained the same solution as before. The solution for a strong coupling ($\nunc\gg\Wcut$) is the same as the one without coupling but with a global attenuation.\\

The asymptotic values observed follow the same  behaviour as we see in Eq. \ref{eq:Asymptotic_Omega_value}. We can define $k_\infty$ as an approximation of the value of $k$ from which this asymptote occurs. An approximation of the value of $k_\infty$ can be given by $k^2$ term dominating the $k$ term in the $\Omega$ term of the equation above. This yields

\begin{equation}
    k_\infty \equiv \dfrac{\kcut}{1-\sin^2\alpha\cos^2\theta} \ . 
\end{equation}

Therefore, with $\alpha=0^\circ$ and $\theta=10^\circ$, the plateau is reached for $k\gg k_\infty$. We can define 
\begin{equation}
    \Omega_{\infty}\equiv\Omega(k\gg k_\infty).
\end{equation}

Even though it is challenging to fully grasp the impact of the coupling just by examining the dispersion relation, we can define $\tilde{\Omega}=\Omega/\phic$, so that this variable satisfies V20 dispersion relation. In particular, we can first look at the asymptotic expressions for large and small values of the coupling parameters:

For small values of $\nunc \ll \Omega$, the term $\phic$ tends towards 1, which means that the solution of our equation approaches $\Omega_{\rm MHD}$; if the coupling is much slower than the development of the instability, the charged fluid undergoes the instability without being influenced by the neutrals and therefore follows the equations of V20. Conversely, when $\nunc \gg \Omega$, the term $\phic$ tends towards $\sqrt{\rhoc/\left(\rhoc+\rhon\right)}$ . The equations are then identical to those of the decoupled case but with an overall attenuation factor: since the coupling is strong, the charges must first drag the neutrals, and thus the gravity $\gc$ is diluted. These two cases can be decomposed into two regimes. Before $\tilde{k} = 1$, the growth follows $k^{1/2}$, while beyond this point, it reaches the previously described asymptotes. In a more general case, at large scales (i.e. small values of $k$), ions are fully coupled to neutrals and, thus, we find a growth as $k^{1/2}$, following the perfect coupling model. As soon as we enter the range where $\nunc \ll \Omega$, friction loses its significance and the growth rate eventually converges to that of the collision-less case. 

\subsubsection{Anisotropy of grown instabilities}

The magnetic field naturally produces anisotropic filamentary structures \citep{1995ApJ...438..763G}. The objective of this section is therefore to quantify this anisotropy in the magnetic RT instability and  the impact of the ambipolar diffusion on the anisotropy.

\begin{figure*}
    \centering
    \includegraphics[width=.98\linewidth]{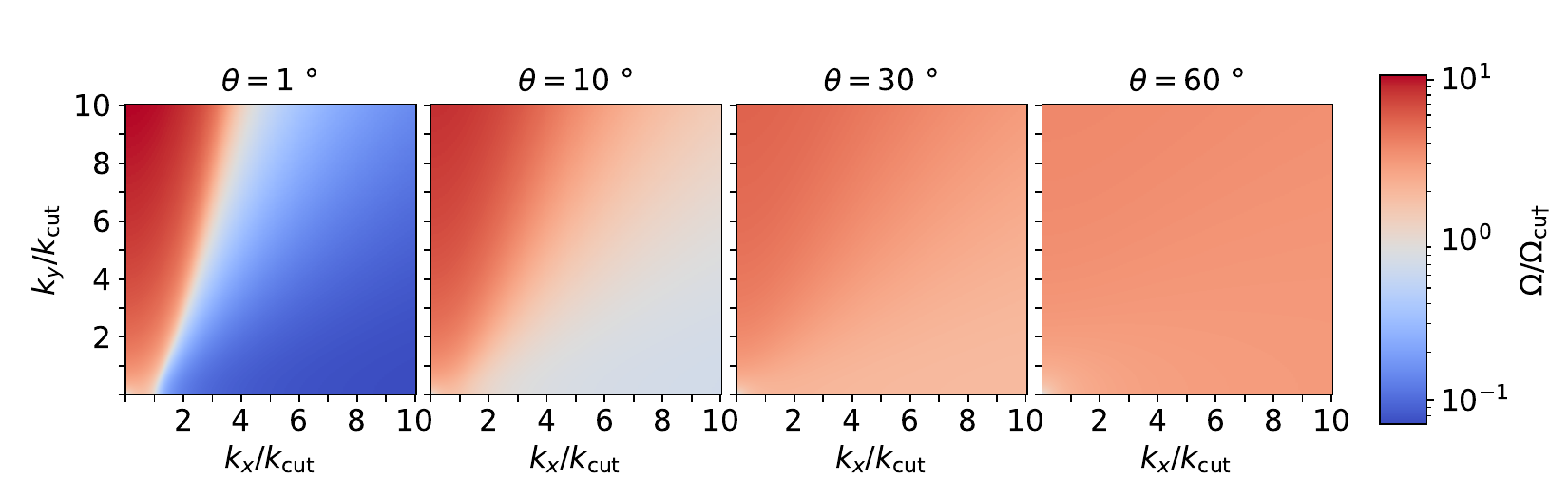}
    \caption{Maps of the value of $\Omega/\Wcut$ without coupling ($\nunc=\nucn=0$) in the $(k_x/\kcut,k_y/\kcut)$ plane. The magnetic field inclination $\theta$ increases from left to right, taking the values $1^\circ, 10^\circ, 30^\circ$, and $60^\circ$.}
    \label{fig:map_kx_ky_mhd}
\end{figure*}

In light of the discussions linked to Eq.~\eqref{eq:dispersion_gn0}, it is important to first recall the behaviour of the instability in the MHD framework.Indeed, the more complex case with coupling appears to be similar to this initial framework but with a few adjustments. In Fig.~ \ref{fig:map_kx_ky_mhd}, we plot the map of $\Omega/\Wcut$ values without coupling in the ($k_x/\kcut,k_y/\kcut$) plane. From left to right, the inclination angle of the magnetic field gradually increases from $1^\circ$ to $60^\circ$.  A clear trend in growth rate anisotropy can be seen when comparing the first and last maps. Indeed, while the last map is almost uniform, the first map shows very large contrast differences. Between these two maps, the contrast decreases uniformly: the more  constraining it becomes, then the larger the magnetic tension. The shape of the map also shows that the situation is not symmetric in $k_x$ and $k_y$: for $k_x/\kcut > 1$, we observe a strong slowdown in the growth rate for small $\theta$, which is not visible for $k_y/\kcut > 1$. \\

To quantify this anisotropy, we define a function $A_\alpha$:
\begin{align}
    A_\alpha &\equiv\dfrac{\Omega(\alpha=90^\circ)-\Bar{\Omega}}{\Omega(\alpha=90^\circ)+\Bar{\Omega}} \,,\\
    \Bar{\Omega}&\equiv\dfrac{2}{\pi}\int_{0}^{\pi/2}\Omega(\alpha)\,\d\alpha \,,
\end{align}
where $\Bar{\Omega}$ is the mean value of $\Omega$ over $\alpha$. Physically, this choice is guided by the understanding that in a given medium, what we observe is the variation or contrast in growth across different directions. By defining the anisotropy and averaging it over all values of $\alpha$, we can capture the contrast differences in each direction and better understand which modes can genuinely develop. When the anisotropy function is close to 1, it indicates strong directional dependence: only specific values of $\alpha$, primarily around $\pi/2$, can effectively develop. This highlights a pronounced anisotropic effect where growth is concentrated in certain directions. On the other hand, an anisotropy value close to 0 implies nearly uniform development in all directions, suggesting that the magnetic field exerts minimal influence on the directional growth patterns. 

\begin{figure}
    \centering
    \includegraphics[width=\linewidth]{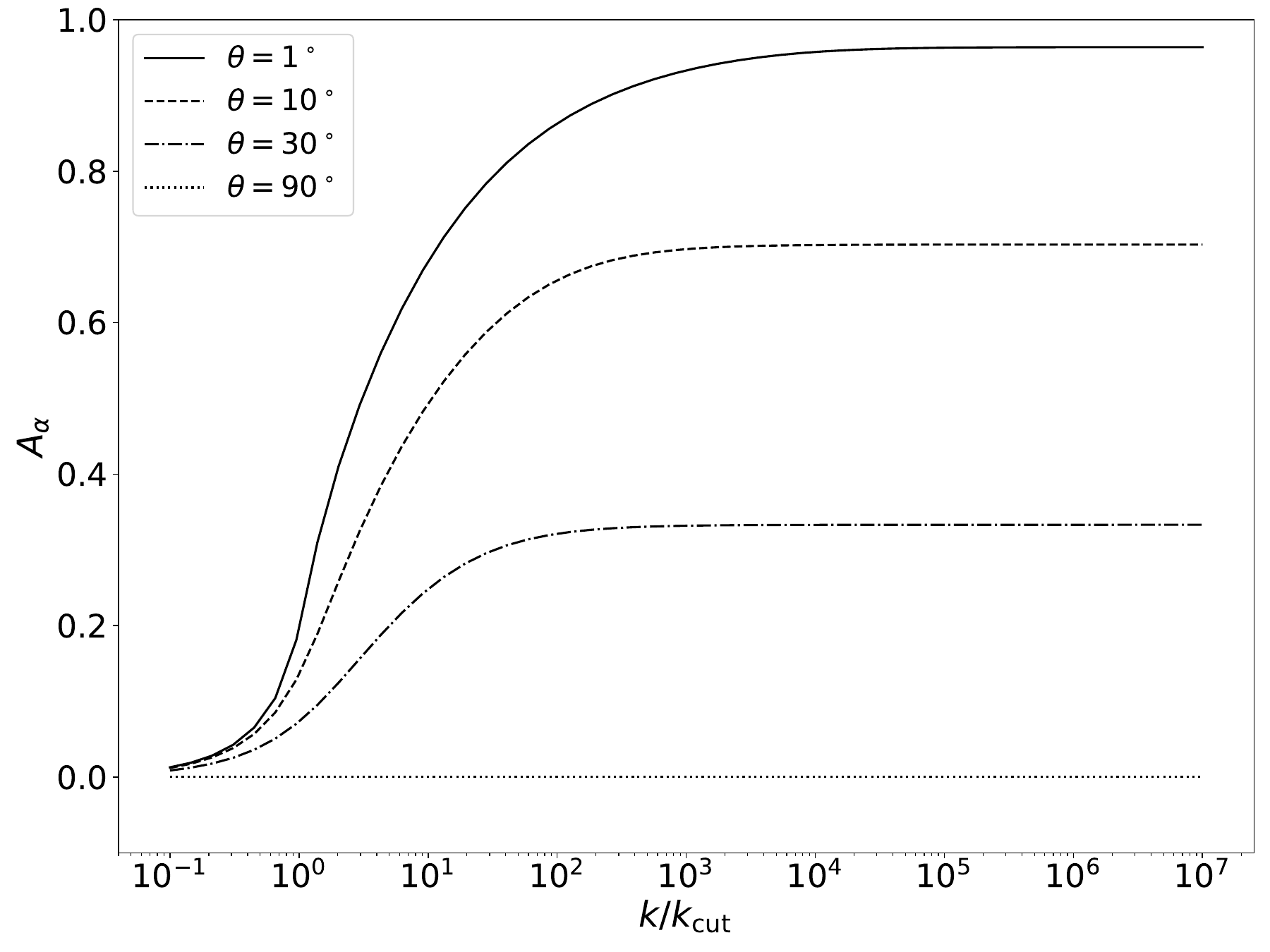}
    \caption{Anisotropy $A_\alpha$ as a function of $k/\kcut$, for various values of the magnetic field inclination $\theta$ (different line style). }
    \label{fig:anis_mhd}
\end{figure}

We plot the anisotropy $A_\alpha$ in Fig.~ \ref{fig:anis_mhd} as function of $k/\kcut$ for various values of $\theta$. The initial observation is the distinct variation in behaviour across angles. For $\theta = \pi/2$, the anisotropy remains consistently at the level of zero, indicating that the magnetic field is unable to influence the RTI in any way other than isotropically. In other words, there is no directional dependence when $\theta$ is perpendicular. In contrast, as $\theta$ drops towards 0, we observe an increase in the anisotropy parameter, reaching its maximum at $\theta = 0$. This confirms that when the magnetic field is aligned with the instability (parallel configuration), it has a pronounced impact on the growth in different directions. The role of $k/\kcut$ is also noteworthy. For values below $k/\kcut = 1$, anisotropy remains almost negligible. However, as $k/\kcut$ increases, the anisotropy progressively rises, eventually approaching an asymptotic value at larger $k/\kcut$ values, indicating that the influence of the magnetic field on directional growth stabilises at high $k/\kcut$.

\begin{figure}
    \centering
    \includegraphics[width=\linewidth]{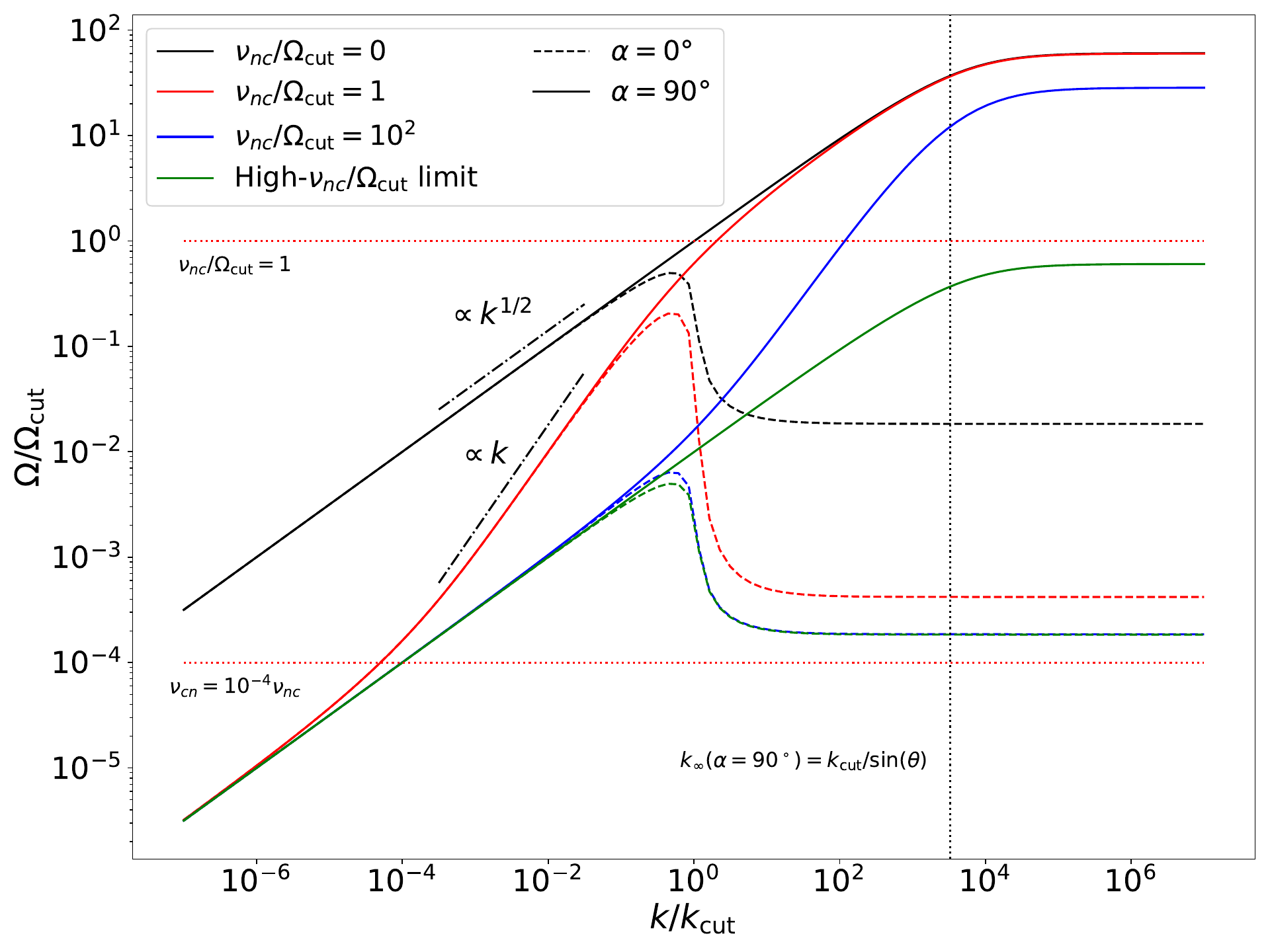}
    \caption{Solutions to the dispersion relation for various values of the coupling parameter $\nunc/\Omega$: 0 (black, MHD limit), 1 (red), $10^2$ (blue), and $\infty$ (green, perfect coupling), for two values of the angle $\alpha$: $90^\circ$ (solid) and $0^\circ$ (dashed). The solutions are normalised by $  \Wcut$. We set $\gn=0$, $\theta=1^\circ$, $d=0.25$, $\rhozeroc/\rhozeron=10^{-4}$ as in Fig.~\ref{fig:omega_vs_k_gn0}. }
    \label{fig:gn0_var_Alpha}
\end{figure}

Next, we  turn to the coupling effect and in Fig.~\ref{fig:gn0_var_Alpha}, we plot the solutions to the dispersion relations for various values of the coupling parameter and various values of the angle $\alpha$. We plot the two extreme cases,  $\nunc/\Omega=0$ (i.e. MHD mode in black) and $\nunc/\Omega=\infty$ (i.e. charged and neutral fluids totally coupled in green). Between these two limits, we can draw two values of the coupling parameter. As  stated above, we can see the impact of the coupling parameter on the form of the curve. Following, for instance, the red curve: at small scales, the charges do not interact with the neutrals, since the growth time of the instability is much shorter than the ion-neutral collision time ($\Omega > \nu_{nc}$). As a result, the coupling cannot be established during the development of the instability. Then, when $\Omega < \nu_{nc}$, the charges start to feel the friction with the neutrals. Finally, at even larger scales, when $\Omega < \nu_{cn}$, the neutrals begin to respond to the charges, and the two fluids behave as a single coupled fluid.. In addition to this behaviour, this figure provides a first insight into the difference in anisotropy: for large values of $\nunc$, the curves are identical up to an attenuation factor, meaning the anisotropy remains the same. Since decoupling occurs at increasingly larger scales as $\nunc$ decreases and given that the anisotropy caused by the magnetic field only appears at small scales, no impact on anisotropy will be observed for small $\nunc$. Between these scales, for the $\nunc$ values plotted, we observe that at $k / \kcut$, the gap between the curve for $\alpha = \pi / 2$ and $\alpha = 0$ is larger than in the MHD case. Therefore, to address the anisotropy, we go on to consider $\nunc/\Wcut = 10$. 

\begin{figure}
    \centering \includegraphics[width=\linewidth]{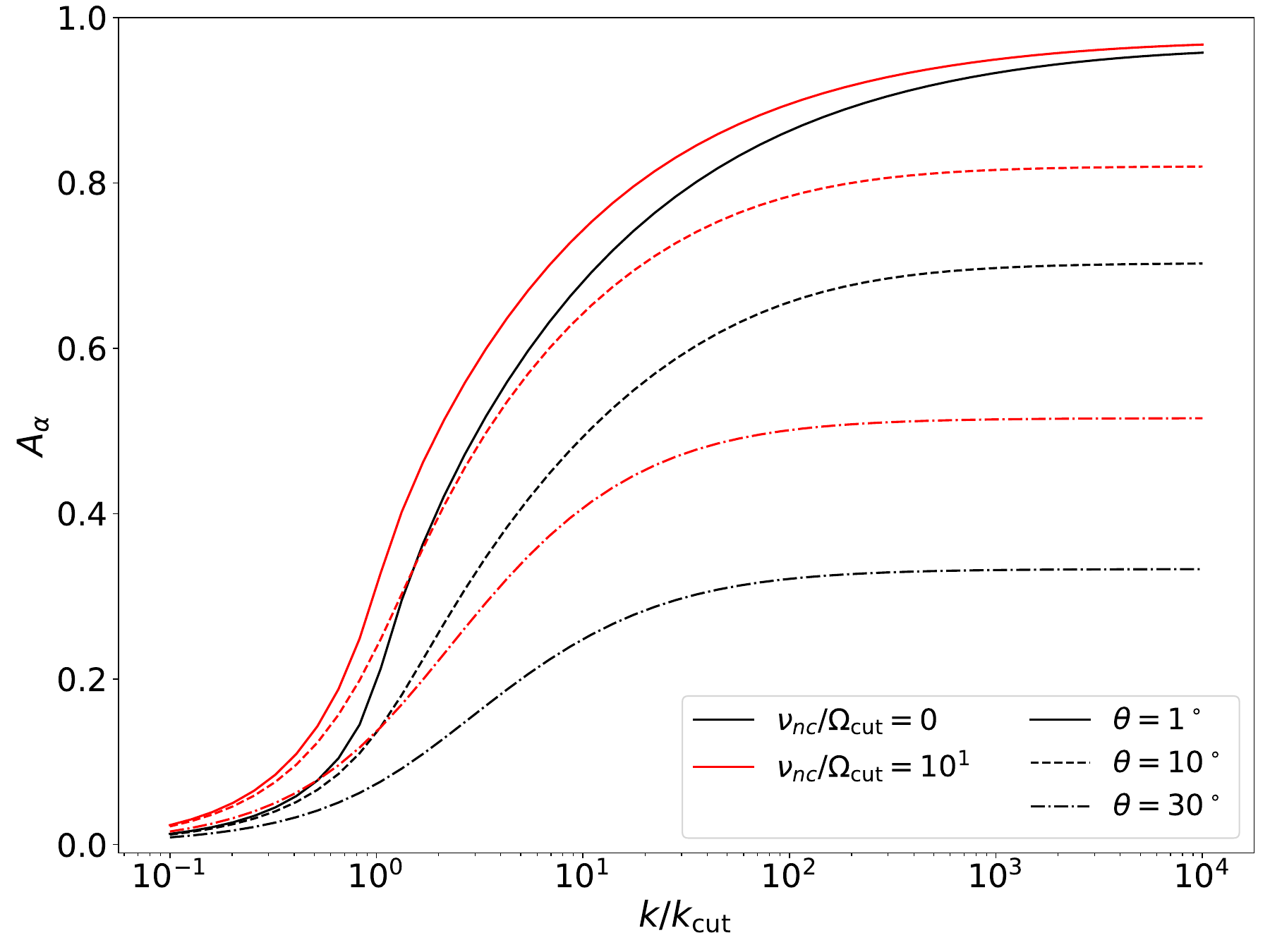} \caption{Variation of the anisotropy as a function of $k/\kcut$, for various values of $\theta$ (plain $\theta=1^\circ$, dashed $\theta=10^\circ$, dash-dotted $\theta=30^\circ$), for $\nunc/\Wcut=0$ (black) and $\nunc/\Wcut=10$ (red).}
    \label{fig:Anis_vs_k}
\end{figure}

We plot in Fig.~\ref{fig:Anis_vs_k} (similarly to what is shown in Fig.~\ref{fig:anis_mhd}), the anisotropy as a function of $k/\kcut$ for different values of $\theta$. We see that regardless of the angle $\theta$ or the value of $k/\kcut$, the anisotropy curve with coupling consistently lies above that of the MHD case. This confirms our initial observations: coupling tends to enhance the anisotropy induced by the magnetic field. The asymptotic behaviour of the anisotropy could be explained by Fig.~\ref{fig:gn0_var_Alpha}: for larger $k/\kcut$, whatever the value of the angle $\alpha$ the growth rate reaches an asymptotic value given by Eq.~\eqref{eq:Asymptotic_Omega_value}, which leads to a plateau for the anisotropy.

\begin{figure}
    \centering
    \includegraphics[width=\linewidth]{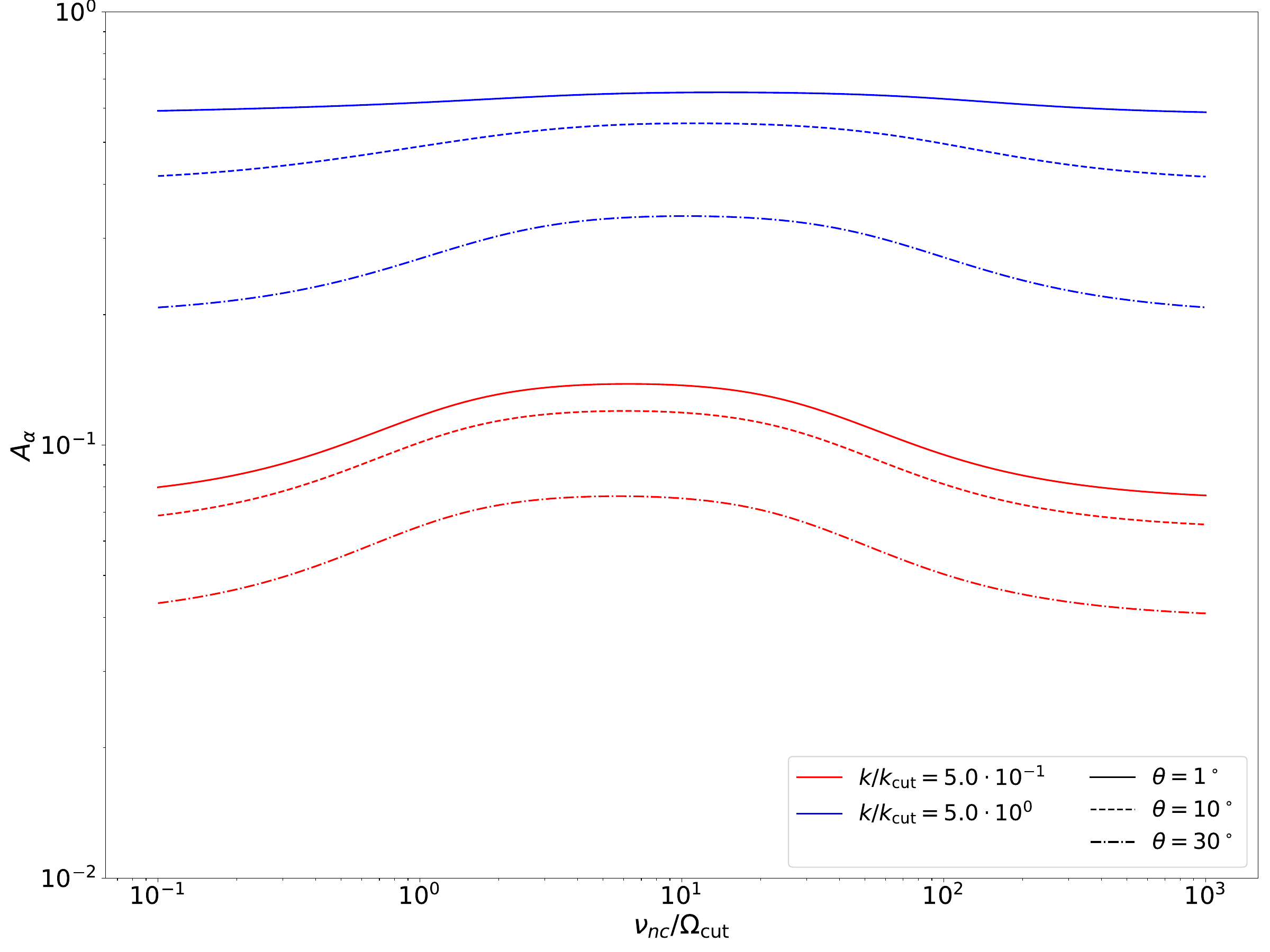}
        \caption{Plot of the anisotropy for various values of  $k/\kcut$ different colours (blue $k/\kcut=5$, red $k/\kcut=0.5$) and for various values of $\theta$ different line styles (plain $\theta=1^\circ$, dashed $\theta=10^\circ$, dash-dotted $\theta=30^\circ$)  as a function of of the coupling parameters. }
      \label{fig:Anis_vs_nu}
\end{figure}

To conclude this analysis of the anisotropy, we plotted the anisotropy as a function of the coupling parameter for various values of $k/\kcut$ and $\theta$ in Fig.~\ref{fig:Anis_vs_nu}. First, we note that the larger the $k/\kcut$, the stronger the anisotropy, which can be anticipated based on the previous figures. Specifically, for high of values $k/\kcut$ with $\theta\simeq 0$, we find a constant anisotropy equal to 1. Conversely, for $k/\kcut$ values below 1, we observe minimal anisotropy. A noteworthy point across this plot is the particular dependence of anisotropy on the coupling parameter. Except for the previously described extreme cases, all curves exhibit a similar behaviour: at extreme values of $\nunc/\Wcut$, they reach flat asymptotes, while within the previously specified range where decoupling and thus ambipolar diffusion occur,  anisotropy reaches a peak. This figure therefore clearly highlights the significant role that ambipolar diffusion can play in enhancing anisotropy.

If we shift from a theoretical framework to a more astrophysical context, we must consider that the magnetic field, the effective gravity, and the coupling strength between neutrals and charges are no longer variables but rather fixed parameters. The range in which anisotropy manifests significantly holds almost independently of the observational scale or the inclination angle of the magnetic field.

\section{The RTI in the magnetised, partially ionised ISM}\label{S:Appli}

\subsection{Context and physical justification of the model}\label{ss: Context_Appli}

The astrophysical context we aim to study here is specifically that of the ISM. We restricted our analysis to the multiphase, diffuse and very weakly ionised ISM. In this context, we considered three species: neutrals (either H or H$_2$), ions (either H$^+$ or C$^+$),  and dust grains.  We therefore aimed to apply the model developed above to this type of medium. The incompressibility assumption makes sense while considering diffuse ISM phases. In atomic (warm and cold neutral) phases that are the subject of our interest here, turbulence properties is deduced from the 21 cm {\sc Hi} line emission \citep{2005ApJ...624..773H}. In the cold neutral medium (CNM), it is found that the turbulent velocity dispersion has a median value of $2.8$ km/s, which implies a moderately supersonic turbulence with sonic Mach numbers $M_{\rm s}\sim 3$. In the warm neutral medium (WNM), higher turbulent velocities induce slightly subsonic to transonic flows because the sound velocity is also higher $c_{\rm s} \sim 10$~km/s \citep{2014AA...567A..16S}. The total magnetic field derived from Zeeman splitting effect in {\sc Hi} emission (WNM) or absorption (CNM) lines have a median values of $B = 6~\mu$G in the CNM, while measurements provide less accurate estimates of $B_{\rm tot} = 5-10~\mu$G in the WNM \citep{2005LNP...664..137H}. We find typical Alfv\'enic Mach numbers are $M_{\rm a} \sim 3$ in the CNM and $\lesssim 1$ in the WNM. We conclude due to these moderate Mach numbers that CNM (but also  likely WNM) show a moderate compressibility. In any case, the compressibility applies mostly at scales close to the turbulence injection scale. As scales of the ambipolar diffusion are concerned they are below the sonic scale where the gas turbulent velocity matches the ambient sound speed. Hence, we are dealing  with a part of the turbulent cascade lying in the incompressible regime (see e.g.\ \citet{Beattie24}).

\subsection{Modelling ions and dust as a single charged fluid}

In the diffuse ISM, we  need to consider a priori a tri-fluid model for our system composed of neutrals, ions, and dust. If we want to use our bi-fluid model and the corresponding results presented in Section \ref{S:Results}, we need to adopt a simpler approach by gathering all the charges into a single charged fluid. This approach is motivated by \citet{Zaqarashvili} who demonstrated how to unify a H$^+$/He$^+$ plasma in a single fluid. The condition is that $\Omega$, the gyrofrequencies of each charged species, is much larger than the collision frequency, $\nu$, between distinct charged species. In term of physics, this means that the charged species are coupled due to a strong magnetic field and not due to collisions. We describe below the different components of this charged fluid (i.e. ions, PAHs, and dust) and show that the conditions are satisfied to gather them into a single charged fluid,  described with the index $c$. \\ 

Dust grains in the diffuse ISM cover a size distribution ranging from subnanometer particles (i.e. polycyclic aromatic hydrocarbons, PAHs) to submicrometer grains \citep{HD23}. Through the competition between electron collection, recombination of ions on the surface of grains, and photoemission of surface electrons, dust grains acquire an electric charge of $q=Ze$ \citep{WD01c}. In the presence of a significant radiation field ($G_0 > 1$), the grain charge is on average positive and increases with the grain radius, $a$. Through the Lorentz force, dust grains of a mass, $m$, are coupled to the magnetic field of intensity, $B$, on a timescale of $\tau_{\rm Lar} = 1/\Omega_{\rm Lar}$, where $\Omega_{\rm Lar} = qB/mc$ is the Larmor frequency and $c$ is the speed of light. Through collisions with gas particles, they are also coupled to the neutral gas on a timescale of $\tau_{\rm drag} = m/(\rhon\sigma c_{\rm s}),$ where $\sigma=\pi a^2$ is the grain cross-section and $c_{\rm s}$ is the sound velocity in the gas. The ratio of these two timescales, the Hall factor $\Gamma = \Omega_{\rm Lar}\tau_{\rm drag}$, expresses in the CNM:
\begin{equation}
    \Gamma^{\rm CNM} \simeq 20\, Z \left(\frac{\nn}{50\,{\rm cm}^{-3}}\right)^{-1}\left(\frac{a}{0.1\,\mu{\rm m}}\right)^{-2}\left(\frac{B}{6\,\mu{\rm G}}\right)\left(\frac{T}{100\,{\rm K}}\right)^{-1/2} \ .
\end{equation}
It is much larger than 1 for all sizes relevant to the diffuse ISM, implying that dust grains are all strongly coupled to the magnetic field and follow on average the dynamic of ions \citep{Guillet07}. We could argue that neutral grains can not be coupled to the magnetic field. This remark is particularly relevant for PAHs because the smaller the grains, the larger the probability for the grain to be electrically neutral. Actually, grain charges tend to fluctuate. Their neutral state is only temporary, before they acquire a charge again. This stochastic process is characterised by a timescale, $\tau_Z$, that is much shorter than $\tau_{\rm drag}$ \citep[their Fig.~1]{Yan2004}. This implies that grains do not have the time to couple to the gas during their neutral phase and remain coupled to the magnetic field whatever the value of their instantaneous charge. It is therefore justified to assume that all dust grains (including PAHs) always remain coupled to the magnetic field whatever their charge state. We will therefore consider ions and dust grains as a single fluid that is 'frozen' in the magnetic field lines.\\

The properties of our charged fluid is the sum of the properties of ions, PAHs, and large ($5 - 250$ nm) dust grains. For the CNM and WNM, its total mass is dominated by dust grains: $\rhod = 6.410^{-3}\rhon$, $\rhoP = 7.10^{-4}\rhon$ \citep{HD23}, and $\rhoi = \ni \mi$ (see Table \ref{tab:data_WNM_CNM} for the values of $\ni$ and $\mi$ in each phase). The collisional cross-section carried by PAHs is six times higher than the cross-section carried by grains \citep{HD23}. Moreover, the electric dipole induced by a charged PAH on a gas particle increases the collisional cross-section and makes it independent of the collisional velocity or PAH size (the Langevin  cross-section $\sigVL=2.1\times10^{-9}~\rm{cm^3/s}$, see \citet{Jean_2009,Brahimi_2020}). The same cross-section applies to ions. Next, we compute the charged-neutral collisional frequency, $\nunc$, as
\begin{align}
     \rhon\nucn & = \rhoc\nunc = \rho_i\nuni+\rhoP\nunP+\rhod\nund \,,\\
     \rhoi\nuni & = \nn\dfrac{\mi\mn}{\mi+\mn}\ni\sigVL\,, \\
     \rhoP\nunP & \simeq \rhon\nP\sigVL\,, \\
    \rhod\nund & \simeq \rhon\langle n\sigma\rangle_{\rm d}c_{\rm s} \,,
\end{align}
where $\nn$ (resp. $\ni$, $\nP$ and $\nd$) is the density of neutral particles (resp. ions, PAHs, and dust grains), $\mn$ (resp. $\mi$) are the mass of neutral particles (resp. ions), and $\langle n\sigma\rangle_{\rm d} = 3.10^{-21}\nH$ \citep{WD01b}. We find the contribution of PAHs and dust grains to be negligible compared to the ion term both for the CNM and WNM and, thus, we can neglect them. We then obtain:\ 
 \begin{align}
     \nucn\simeq & \frac{\rhoi}{\rhon}\nuni \simeq \dfrac{\rhoi}{\mi+\mn}\langle\sigma_{\rm cn}v\rangle\,,\\
      \nunc =  & \frac{\rhon}{\rhoi+\rhoP+\rhod}\nucn\,.
 \end{align}

\subsection{Effective gravity due to radiation pressure}

In the diffuse ISM, we can a priori assume that the gravity of the surrounding stars, as well as self-gravity, can be ignored. Several mechanisms could justify the appearance of a RTI by creating an effective gravity. The one we aim to explore relies on radiation pressure. We consider the following framework: a star irradiates an ISM cloud. In doing so, the cloud will be subject to radiation. Radiation can first push the neutrals and simply ionise them,  leading to them being absorbed. Beyond a distance where all this radiation is absorbed, we can assume that the neutrals will no longer experience any radiation pressure (see chapters  6-7 in \citet{2006agna.book.....O}). The remaining radiation pressure will therefore primarily affect dust particles. \citet{Jacquet_2011} demonstrated that  this pressure generates a steady acceleration of the medium $\textbf{a}$ that, through the equivalence principle, is equivalent to a gravity field of $\textbf g=-\textbf{a}$ exerting its effects in an inertial frame,  which can trigger an RT instability. \citet{2014MNRAS.443..547P} also explored this in the case of the Crab Nebula, where supernova ejecta formed a shell subjected to a matter wind from the pulsar, enabling the appearance of an RTI.

From the energy conservation, the total energy emitted by dust is equal to the total starlight energy absorbed by dust. In the neighbourhood of a bright star, the interstellar radiation field is highly anisotropic. The transfer rate of momentum by radiation pressure acting on dust is therefore proportional to the total power, $\mathcal{L_{\rm d}}$, emitted by dust per H (in erg/s/H), here computed for the interstellar radiation field (ISRF, $G_0 = 1$). The same is true for PAHs ($\mathcal{R}_{\rm P})$, but not for ions which do not absorb starlight. The \cite{HD23} diffuse dust model gives $\mathcal{L}_{\rm d} = 3.4\times10^{-24}$\,erg/s/H and $\mathcal{L}_{\rm P} = 1.6\times10^{-24}$\,erg/s/H. The effective gravitational field $\gc$ for the charged fluid is then computed from the conservation of its momentum:
\begin{equation}
    \rhoc\gc  = G_0\frac{\left(\mathcal{L}_{\rm P}+\mathcal{L}_{\rm d}\right)\nH}{c}\,.
\end{equation}

We find 
\begin{equation}
\gc \simeq 5\times10^{-9}\,G_0\,.
\label{eq:gc_G0}
\end{equation}

\subsection{Application to the CNM and WNM}
\begin{table}
    \centering
     \caption{Values of the parameters in the warm neutral medium (WNM) and cold neutral medium (CNM)}
    \begin{tabular}{c|c|c}
  
    \hline
        Phase & WNM & CNM  \\
        \hline
          $B_0\,[\rm G]$ &$5\cdot10^{-6}$ &$6\cdot10^{-6}$\\
          $\nn\,[{\rm cm}^{-3}]$&$0.2$&$50$\\
          $\ni\,[{\rm cm}^{-3}]$&$7\cdot10^{-3}$&$2.4\cdot10^{-2}$\\
          Ions& H$^+$&C$^+$\\
          Neutral& H$+$He&H$+$He\\
          $\mi\,[m_{\rm H}]$&1&12\\
          $\mn\,[m_{\rm H}]$&1.21&1.21\\
          $\bar{T}(K)$ & 8000 K & 50 K \\
          $\cac \, [{\rm cm.s}^{-1}]$&$1.2\cdot10^7$ &$1.55\cdot 10^6®$ \\
          \hline
    \end{tabular}
    \tablefoot{ $B_0$ is the magnetic field strength, $\nn$ (resp. $\nc$) is the density in the neutral (resp. charged) fluid. The temperatures are mean values. \citep{Brahimi_2020}. }
    \label{tab:data_WNM_CNM}
\end{table}

Using Table \ref{tab:data_WNM_CNM} and Eq.~\ref{eq:kcut}, we can estimate the cutoff wavelength in each phase:

\begin{align}
    L_{\rm cut}^{\rm CNM} \equiv 2\pi/\kcut^{\rm CNM} \simeq \frac{0.5}{G_0}\,\text{kpc}\,,\\
    L_{\rm cut}^{\rm WNM} \equiv 2\pi/\kcut^{\rm WNM} \simeq \frac{100}{G_0}\,\text{kpc}\,.
\end{align}
  
It then appears that this length for the CNM is typically on the order of tens of parsecs for a $G_0$ value of $10$. With these considerations, the only instabilities that can grow are those described by the asymptotic behaviour (i.e for $k\gg k_{\infty}$). \\

Similarly, we can estimate the ratio of the collisional frequency between the neutral and charged fluids to the cut-off frequency of RT instability (Eq.~\ref{eq:wcut}) as:\ 
\begin{align}
 \left(\nunc/\Wcut\right)^{\rm CNM} & =\frac{1.1\cdot10^6}{G_0} \ , \\
  \left(\nunc/\Wcut\right)^{\rm WNM} & = \frac{6.0\cdot10^4}{G_0}\ . 
\end{align}

It thus becomes clear that unless there is an extremely high radiation pressure, coupling will always occur on timescales much shorter than those over which the RT instability could develop. As a result, with respect to this instability, the two fluids will behave as if they are perfectly coupled. Therefore, we cannot expect any observable signature of ambipolar diffusion in the development of this instability: what could be observed is merely strong scale-independent anisotropy.\\

 We can go on to estimate the growth time of the instability:
 \begin{align}
 t^{\rm CNM}&  \equiv \frac{1}{\Omega^{\rm CNM}_\infty\left(\alpha=\frac{\pi}{2}\right)}\simeq \frac{1.4}{G_0}\,\text{Myr} \ ,  \\
 t^{\rm WNM}&  \equiv \frac{1}{\Omega^{\rm WNM}_\infty\left(\alpha=\frac{\pi}{2}\right)}\simeq \frac{18}{G_0}\,\text{Myr} \ . 
\end{align}
With realistic values (for $G_0$ (1-100)), we expect growth times on the order of, at best, $0.1$ Myr .
To better grasp the order of magnitude of this growth time, we applied it to the case of the environment of the star Merope, a star in the Pleiades, which is surrounded by an interstellar medium cloud from the Taurus region. The interaction between these two systems has been documented by \citep{Gibson_2003}. We can thus recall that the Pleiades came into contact with this cloud at \( \tau \approx 0.1 \, \mathrm{Myr} \). Moreover, the radiation field from Merope  can be estimated as \( G_0 = 20 \) . With this radiation field, we obtain a growth time for the instability comparable to the characteristic interaction time with the star. Therefore, the instability could, in principle, develop. Two limitations must be taken into account: on the one hand, the calculated time corresponds to the fastest-growing mode ($\alpha=90^\circ$), meaning that few other modes (with different angle values $\alpha$) have sufficient time to fully develop; on the other hand, since the two timescales are comparable, the actual development of this mode remains rather marginal.\\

The conclusion drawn here is, at first glance, rather negative. It is therefore important to recall that the analysis presented here is a linear analysis. However, as pointed out by \citet{10.3389/fspas.2022.789083}, ambipolar diffusion can play a significant role when the instability enters a nonlinear regime with large-scale effects. Indeed, \citet{refId0} found, using a 2.5D numerical simulation, that the growth rate of small-scale modes during the nonlinear phase can be significantly higher with ambipolar diffusion than without. Other, more recent works also support this view \citep{Braileanu_2021a,Braileanu_2021b}. These results highlight the non-negligible impact of ambipolar diffusion in the structuring of the instability during the nonlinear phase.

\section{Conclusion}\label{S:Concl}

We  analytically studied the effect of ambipolar diffusion on a MRTI for an oblique magnetic field. We used a two-fluid incompressible formalism. To study an instability triggered by radiation pressure rather than gravity, we chose to differentiate the acceleration imparted on each species. This approach leads to the following conclusions:

\begin{enumerate}

\item  In the presence of acceleration on both species: in this case, the dispersion relation yields two modes as solutions: a modified HD mode and a modified MHD mode the behaviours of the two fluids will be linear combinations of these two modes. These two modes are defined such that in the asymptotic case, where the coupling is zero, they correspond to the HD and MHD growth rates, respectively.

\item When gravity acts only on the charged species, the system presents a single solution for the growth rate of the instability. This solution transitions from an uncoupled solution to a fully coupled one, which is simply the uncoupled solution attenuated by a global factor. Between these two asymptotic solutions, the signature of the ambipolar diffusion is a range in length within which the growth rate changes slop.

\end{enumerate}

Applying the results of the dispersion relation to the study of the interstellar medium, we showed that the magnetic field tends to enhance the anisotropy in the structuring of the RTI, especially when the field is tangent to the interface. We also demonstrated that this anisotropy becomes stronger with coupling, but only for intermediate coupling values.\\

We developed a representative framework for the interstellar medium by calculating an effective gravity that could be provided by the radiation pressure of a star located behind an interstellar cloud on our line of sight. This calculation revealed that the cutoff length below which the instability reaches its asymptote for all wavelengths is very large. Consequently, the anisotropy developed by the instability is then very strong but independent of the scale, as it saturates below a certain length scale. Under these parameters, the RTI develops over a timescale significantly longer than the characteristic time of complete coupling between neutrals and ions. Therefore, there is no clear signature of ambipolar diffusion in the growth of the RTI for the estimated values. However, it is conceivable that  stars with stronger radiation or another force acting as a constant anisotropic acceleration could allow for the development of the RTI  within the framework we  describe in this article.\\

This work is a linear analysis of the RTI that does not capture the full complexity of the  actual behaviour behind the instability. In effect, a nonlinear study could be of great interest, as partial ionisation may have a much more significant impact when nonlinear effects come into play \citep{10.3389/fspas.2022.789083}. This remark will be the basis for a forthcoming numerical study.

\begin{acknowledgements}
The authors are grateful to O. Berné, F. Boulanger, K. Ferrière for fruitful discussions. We also thank A. Cogez for his contribution during his internship. This work was supported by the Thematic Action “Programme National Physique Stellaire” (PNPS) of INSU Programme National “Astro”, with contributions from CNRS Physique \& CNRS Chimie, CEA, and CNES. 
\end{acknowledgements}

\end{document}